\pdfoutput=1
\documentclass[runningheads]{llncs}
\usepackage[T1]{fontenc}

\usepackage{newtxtext}
\usepackage[varvw]{newtxmath}

\usepackage{graphicx}
\usepackage{xcolor}
\usepackage{hyperref}
\hypersetup{hidelinks}

\usepackage{booktabs}
\usepackage{xspace}
\usepackage[nohyperlinks]{acronym}
\usepackage{mdframed}
\newcommand{\name}{\textsc{puAMM}\xspace}
\newcommand{\R}{\mathbb{R}}
\newcommand{\E}{\mathbb{E}}

\newcommand{\PX}{\mathbb{P}_X}
\newcommand{\CVaR}{\operatorname{CVaR}}
\newcommand{\VaR}{\operatorname{VaR}}
\newcommand{\TV}{\operatorname{TV}}

\acrodef{AMM}{automated market maker}
\acrodefplural{AMM}[AMMs]{automated market makers}
\acrodef{CFMM}{constant function market maker}
\acrodefplural{CFMM}[CFMMs]{constant function market makers}
\acrodef{CVaR}{conditional value-at-risk}
\acrodef{LMSR}{logarithmic market scoring rule}
\acrodef{LP}{liquidity provider}
\acrodefplural{LP}[LPs]{liquidity providers}
\acrodef{CLOB}{central limit order book}
\acrodef{APR}{annualized percentage rate}
\acrodef{TV}{total variation}
\acrodef{TWAP}{time-weighted average price}

\mdfdefinestyle{pumbox}{%
  linewidth=0.7pt, linecolor=blue!55!black, backgroundcolor=blue!5,%
  innertopmargin=5pt, innerbottommargin=5pt,%
  innerleftmargin=8pt, innerrightmargin=8pt,%
  skipabove=6pt, skipbelow=6pt, nobreak=true}
\surroundwithmdframed[style=pumbox]{problem}
\begin{document}
\title{A Portfolio Underwriting AMM for Tail Liquidity in Rare-Event Prediction Markets}
\titlerunning{A Portfolio Underwriting AMM for Tail Liquidity}
\author{Weilin Li\inst{1} \and Stefanos Chaliasos\inst{2,3} \and Jiahua Xu\inst{1}}
\authorrunning{W. Li et al.}
\institute{University College London, London, UK \and National and Kapodistrian University of Athens, Athens, Greece \and zkSecurity}
\maketitle
\begin{abstract}
Prediction market prices are read as probability forecasts, but for rare events they are expensive to back: selling a claim on a low-probability outcome locks nearly its full payout as collateral, so block-sized depth is scarce. Pooling the capital alone does not solve this: a price-bounded, never-default provider serves at most $O(K)$ claims per reserve $K$. To lift this ceiling, we introduce \name{}, an AMM that sells rare-event claims on $n$ markets and underwrites their aggregate liability with one shared reserve, given a joint distribution of outcomes. Two CVaR protections define the mechanism: premiums must cover the average payout over the worst $\sigma_2$-fraction of outcomes, and premiums plus reserve must cover it over the rarer $\sigma_1$-fraction, while traders bear a proportional haircut with probability at most $\sigma_1$. It is provably the frontier of the two protections, and it certifies $\Omega(\sqrt n)\,K$ notional per reserve $K$ across $n$ independent rare-event markets, with markup vanishing as $n$ grows. Numerical experiments show that at $(n,p)=(100,0.02)$ one reserve dollar backs $55.34$ dollars of claims, $54.2$ times the full-collateral ceiling, and that leverage tracks how rarely rare events coincide, not merely how rare they are. On Polymarket order books outside the most liquid quartile, where 10K shares cost $6.6\times$ the mid, \name{} is the cheaper or only venue in $64\%$ of markets at 10K shares and in nearly every market at 100K. A replay on $907$ rare-event markets is consistent with independence and shows no insolvency.

\end{abstract}
\acresetall
\begin{figure}[htb]
\centering
\includegraphics[width=.81\textwidth]{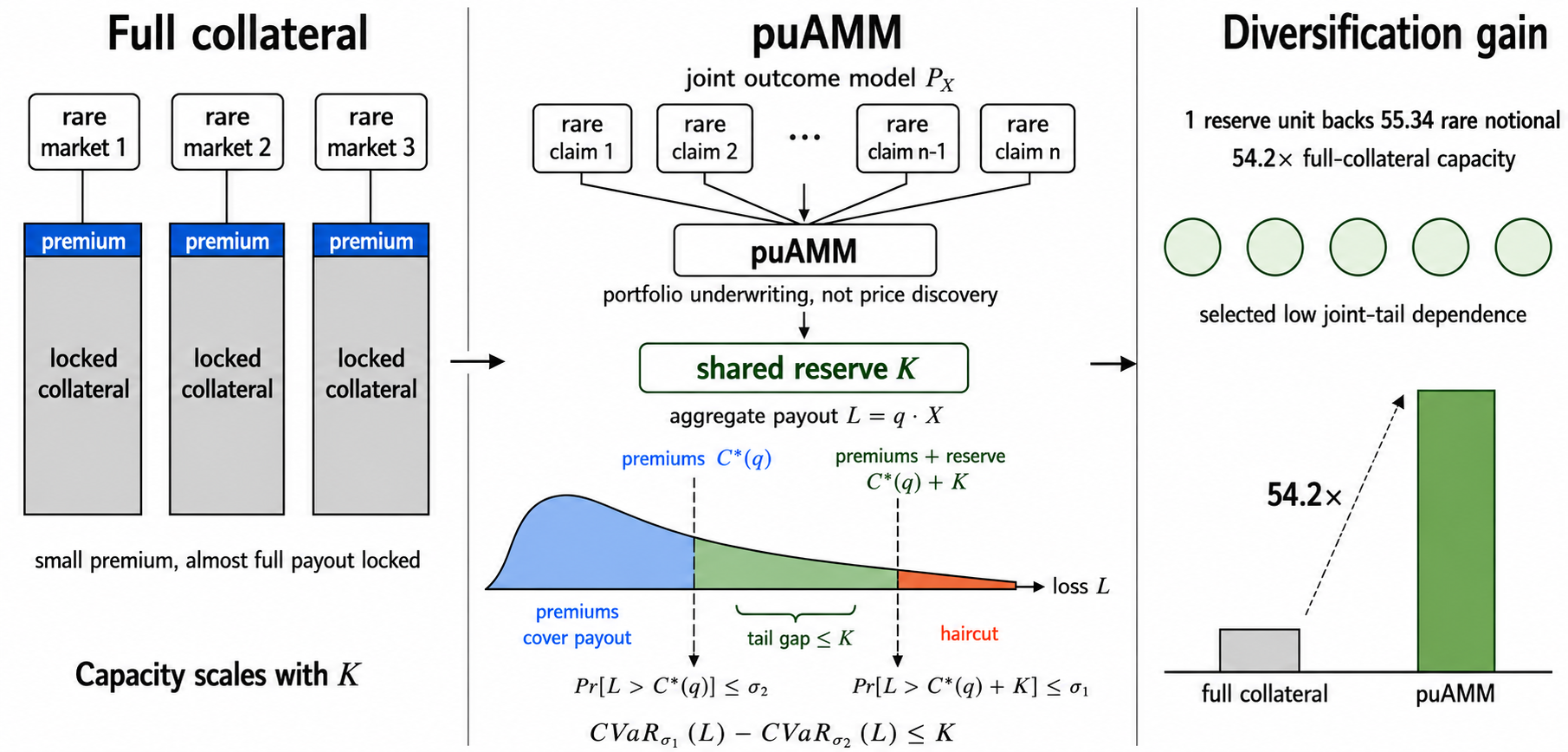}
\caption{\textbf{Overview.} Left: full collateral locks nearly one unit per claim, so capacity is $O(K)$. Middle: under $\PX$, \name{} funds the $\sigma_2$ tail of the payout $q\cdot X$ with premiums and the $\sigma_1$--$\sigma_2$ tail gap with the reserve. Right: one reserve unit backs $55.34$ units of rare-event notional at $(n,p)=(100,0.02)$.}
\label{fig:overview-teaser}
\end{figure}

\section{Introduction}
\label{sec:introduction}

Prediction markets trade claims on future events, and their prices are read as probability forecasts. Many markets are heavily skewed: one outcome, a catastrophe, a default, an extreme political upset, settles with probability of a few percent. Supplying that side is capital-inefficient: selling the claim locks nearly full payout as collateral until resolution (a claim sold at \$0.05 locks \$0.95 per \$1 payout), and on long-horizon markets the lockup commands a carry premium~\cite{page2013calibrated}. As a result, tail liquidity is thin and expensive. On Polymarket, one of the largest venues, 93\% of the active markets with one side priced below five cents cannot fill a 100K-share order on that side below twice the mid (\S~\ref{sec:numerics}), and low-probability contracts trade at wider spreads~\cite{dubach2026anatomy} and, in several market categories, above the frequency with which they pay out~\cite{cardozo2026longshot}. Prediction market AMMs such as pm-AMM~\cite{moallemi2024pmamm} even deliberately allocate less liquidity at tail prices.

This missing depth is the bottleneck we study, and it is an underwriting problem, not a price-discovery one: an accurate quote is only a number until a seller locks collateral behind it. Collateral is siloed today: capital backing one rare event cannot support another, even when the two seldom strike together. Sharing one reserve across markets is the obvious remedy, but pooling alone does not create leverage. A provider who always pays in full must withstand the outcome in which every sold market resolves to the rare event. Whenever unit prices stay bounded away from the full payout, total notional is capped at $O(K)$ per reserve $K$, however the capital is arranged (Proposition~\ref{prop:barrier}). Leverage beyond this ceiling requires a small, controlled chance of shortfall. The question is then: given a tolerated tail risk, how many claims can one reserve back, and at what premium?

We answer this question with \name, a portfolio underwriting \ac{AMM} that serves as a tail-liquidity layer alongside primary venues (Figure~\ref{fig:overview-teaser}). It sells single-market rare-event claims while underwriting their aggregate liability against one shared reserve $K$, taking a joint distribution of outcomes as an input on $n$ listed markets. Traders buy any listed market, alone or in bundles, never parlays. Two \ac{CVaR} protections at tail probabilities $\sigma_1<\sigma_2$ and a least-premium principle pin down the rule. (i) \textit{\ac{LP} non-loss}: premiums alone absorb ordinary payouts, so the least premium covers the average payout over the worst $\sigma_2$-fraction of outcomes. (ii) \textit{Solvency}: premium plus reserve cover the average payout over the rarer $\sigma_1$-fraction. \name{} refuses trades that would violate either protection, and settles a shortfall, which occurs with probability at most $\sigma_1$, by a proportional haircut.

\noindent\textbf{Contributions.} \emph{We derive \name{} from two protection requirements: }\ac{LP} non-loss forces the premium floor and solvency forces the acceptance rule. The pair is exactly the frontier of the two protections: charging less is infeasible and accepting more forces overcharging (Theorem~\ref{thm:frontier}).
The novelty is that the classic \ac{CVaR} does two jobs, pricing and a stricter solvency gate, which yields a refusal rule that always-quoting \acp{AMM} lack.

\emph{Capital utilization and robustness.} We prove how much tail liquidity one reserve can back in \name{}: the maximum notional per unit of reserve across $n$ independent rare-event markets is $\Omega(\sqrt n)$, with a per-unit markup that vanishes as $n$ grows (Corollary~\ref{cor:iid-symmetric}). It does not depend on $K$ or on the order in which trades arrive (Theorem~\ref{thm:ratio-form}, Corollary~\ref{cor:capacity}). A small error of $\epsilon$ in the input joint probability, measured in total variation, changes this capacity only by $O(\epsilon)$ and raises the default probability by at most $\epsilon$ (Theorem~\ref{thm:tv-robustness}, Corollary~\ref{cor:default-robust}). We also show when traders prefer \name{} to the primary venue (Proposition~\ref{prop:block-cutoff}) and what \acp{LP} expect to earn (Proposition~\ref{prop:lp-apr}).

\emph{Numerical and empirical evaluation.} We show that at $(n,p)=(100,0.02)$ one reserve unit backs $55.34$ units of rare-event notional, $54.2\times$ the full-collateral ceiling. The leverage comes from how rarely rare events coincide, not merely from how rare they are. On Polymarket order books, outside the most liquid quartile of markets, \name{} is the cheaper or the only venue for block-sized long-shot demand, and on $907$ resolved rare-event markets tail resolutions are consistent with independence (\S~\ref{sec:numerics}).

\section{Background and Related Work}
\label{sec:related}

\emph{Prediction markets and their \acp{AMM}.} Prediction market prices aggregate dispersed information into probability forecasts \cite{wolfers2004prediction,arrow2008promise}. On Polymarket, one of the largest venues, claims are minted in YES/NO pairs against one dollar of collateral and trade on an order book \cite{rahman2025sok}. Empirical studies show the cost of trading low-probability contracts \cite{dubach2026anatomy,cardozo2026longshot}. The canonical prediction market \acp{AMM} are market scoring rules, notably \ac{LMSR}~\cite{hanson2003combinatorial,hanson2007lmsr}. Decentralized designs add financial structure: liquidity-based sports books stay fully collateralized \cite{amini2025decentralized}, pm-AMM reallocates liquidity across prices \cite{moallemi2024pmamm}, and a partially active \ac{CFMM}, suggested for prediction markets, throttles per-block liquidity within one pool \cite{ko2026partially}. All of them back every claim in full, whereas \name{} shares one reserve across markets.

\emph{Risk-measure market making.} Risk measures enter market making as homogeneous measures in liquidity-sensitive market makers \cite{othman2011liquidity}, convex and risk-measure cost functions \cite{agrawal2011unified,abernethy2013efficient,abernethy2015risk}, utility-indifference pricing \cite{chen2007utility}, and coherent acceptance sets \cite{artzner1999coherent,follmer2016stochastic}. Our \ac{CVaR} usage follows \cite{rockafellar2000optimization,nemirovski2006convex} and risk-management practice \cite{mcneil2015quantitative}. These works use the risk measure to price trades passively, whereas \ac{CVaR} here is the design object, used for least-premium pricing and for a solvency gate that forces refusal and a capacity theory.

\emph{Optimal \ac{AMM} design and DeFi mechanism design.} A parallel literature derives \acp{AMM} from explicit objectives \cite{goyal2023finding,milionis2024myersonian,curry2025optimal} or axioms \cite{frongillo2024axiomatic,angeris2024geometry}, and we follow its \emph{derive-rather-than-posit} style. Optimal curve design \cite{goyal2023finding} maximizes capital efficiency within one fully collateralized market. Profit-maximizing designs \cite{milionis2024myersonian,curry2025optimal} refuse trades in a no-trade region and impose no capital constraint. Combining \ac{AMM} pools \cite{angeris2024geometry} adds up their reserves, but each pool must stay solvent on its own. \name{} instead shares one reserve across markets and refuses trades only to stay solvent.

\emph{Combinatorial and parlay designs} trade on joint outcomes~\cite{hossain2025combinatorial,rana2026parlaymarket}. \name{} only sells single-market claims and uses the joint law only to price and cap its liability. ParlayMarket \cite{rana2026parlaymarket} pools liquidity and information in a learned LMSR-style model and uses \ac{CVaR} only for evaluation, whereas \name{} imposes \ac{CVaR} as a hard constraint.

\section{Model and Design Problem}
\label{sec:model}

\subsection{Markets and the AMM}

There are $n$ binary prediction markets indexed by $i \in [n]$. Market $i$ resolves to $X_i \in \{0,1\}$ ($1$ means YES). Let $X=(X_1,\ldots,X_n)$, $\PX$ be its joint law, and $p_i := \PX[X_i=1]$.

The \ac{AMM} sells single-market YES claims. A unit claim on market $i$ pays $X_i$ at resolution (if not haircut). Its state is a vector $q=(q_1,\ldots,q_n) \in \R_{\geq 0}^n$, where $q_i$ is the total notional of YES claims sold on market $i$. An \ac{LP} capital reserve $K>0$ backs the \ac{AMM}, and the initial state is $q=\mathbf{0}$. An \ac{AMM} is a pair $(C,\tau)$. The cost function $C:\R_{\geq 0}^n \to \R$ gives the cumulative premium collected to reach state $q$, so trades are priced by increments of $C$. The acceptance rule $\tau:\R_{\geq 0}^n \to \{0,1\}$ decides whether a proposed next state is admissible, depending only on that state. So the reserve enters through an acceptance gate, while the standard cost-function interface is preserved.

We restrict to convex path-independent cost functions with $C(0)=0$ \cite{abernethy2013efficient,frongillo2024axiomatic} whose increments are valid prices for binary claims: for every $q\in\R_{\geq 0}^n$, $i\in[n]$, and $\delta>0$,
\begin{equation}
0 \leq C(q+\delta e_i)-C(q) \leq \delta .
\label{eq:valid-price}
\end{equation}

A request for $\delta$ units of YES on market $i$ at state $q$ proposes the next state $q+\delta e_i$ ($e_i$ the $i$-th unit vector) and executes at premium $C(q+\delta e_i)-C(q)$ if $\tau(q+\delta e_i)=1$. A bundle request proposes $q+\Delta$ for $\Delta\in\R_{\geq 0}^n$, priced and gated alike.
Since reaching state $q$ collects premium $C(q)$, the random gross payout and terminal reserve wealth at $q$ are
\[
      L(q) := q \cdot X = \sum_{i=1}^n q_i X_i, \qquad W(q,X) := C(q)+K-L(q).
\]

We make three assumptions.
\textit{Assumption 1 (Single horizon).}
We analyze one underwriting horizon and settle all claims at its end, so any haircut is proportional (\S~\ref{sec:basic-properties}). Every probability bound still holds when markets resolve at different dates, because terminal reserve wealth depends only on the total payout $L(q)$. Reusing capacity freed by early resolutions is left to a dynamic extension, and heterogeneous horizons enter only through the primary venue's carry premium (\S~\ref{sec:application}).

\noindent\textit{Assumption 2 (Distributional input).}
\label{asm:dist-input}
The \ac{AMM} takes an externally supplied joint law $\PX$ of $X$ as input, with no independence imposed. Cost-function market makers and optimal-design frameworks already take the designer's beliefs as input through their initial quotes or priors \cite{hanson2007lmsr,chen2007utility,goyal2023finding,milionis2024myersonian,curry2025optimal}. The addition here is a joint law over event outcomes. Section~\ref{sec:threat} discusses sourcing $\PX$, and \S~\ref{sec:numerics} tests an independent input against $907$ resolved markets.

\noindent\textit{Assumption 3 (Scope and orientation).}
\label{asm:scope}
Each market's low-probability outcome is labeled YES, relabeling if needed. The \ac{AMM} only sells YES exposure ($q_i\geq0$), never buying it back, and $K$ is fixed over the horizon, excluding \ac{LP} entry, exit, and share pricing.

\begin{problem}[Design problem]
\label{prob:design}
Given a joint law $\PX$ and reserve $K$, design a cost function $C$ and acceptance rule $\tau$ for single-market claims that support large tail-side notional per unit reserve, keep insolvency ($W(q,X)<0$) and \ac{LP} capital loss ($W(q,X)<K$) rare at every accepted state, and avoid arbitrary overcharging.
\end{problem}

\subsection{Baseline: The Full-Collateral Capacity Barrier}
\label{sec:baseline}

The simplest tail-liquidity layer is market-by-market limit orders: a provider splits capital $K$ into budgets $K_1,\dots,K_n$ and supplies YES liquidity on market $i$ at unit price $\pi_i$ by bidding for NO at $1-\pi_i$, so each unit sold locks $1-\pi_i$ of worst-case collateral and budget $K_i$ supports only $q_i\leq K_i/(1-\pi_i)$ units. Pooling the budgets lets idle capital follow demand, at the price of concentrated exposures that call for portfolio-level risk control. Pooling alone, however, buys no capacity, because a provider who always pays in full faces an $O(K)$ ceiling under any arrangement.
\looseness=-1
\begin{proposition}[Capacity barrier]
\label{prop:barrier}
Call a state reachable if accepted trades lead to it from $q=\mathbf{0}$. Take any never-default mechanism selling YES claims, that is, $W(q,X)\geq0$ holds $\PX$-almost surely at every reachable state. Let $q$ be reachable with the support condition $\PX[X_i=1\text{ for all }i\in\operatorname{supp}(q)]>0$, where $\operatorname{supp}(q):=\{i:q_i>0\}$. If \(C(q)\leq\sum_i\pi_i q_i\) for some constants $\pi_i\in[0,1]$, then $\sum_i(1-\pi_i)q_i\leq K$, so each claim on market $i$ ties up at least $1-\pi_i$ of the reserve. If every $\pi_i\leq\bar\pi<1$, then $\mathbf{1}^\top q\leq K/(1-\bar\pi)$, an $O(K)$ ceiling. Proof in Appendix~\ref{app:proofs}.
\end{proposition}

\noindent The proof uses only the outcome in which every sold market resolves YES, so pooling does not lift the barrier, and under independence with $p_i\in(0,1)$ the support condition holds automatically. The bound covers pooled liquidity-bounded-loss \acp{AMM} \cite{amini2025decentralized} and belief-optimal full-collateral curves \cite{goyal2023finding} whenever quotes stay bounded away from one. A bounded-loss \ac{LMSR} \cite{hanson2007lmsr} escapes it only by letting quoted prices approach one.

\section{Portfolio Underwriting AMM}
\label{sec:portfolio-underwriting}

\subsection{Protection Requirements and the Least Premium}

\textit{Protection requirements.} For $0<\sigma_1<\sigma_2<1$, an accepted state $q$ must satisfy
\par\vspace{5pt}\noindent
\begin{minipage}{.5\linewidth}
\begin{equation}\label{P1}
\PX[L(q)>C(q)+K]\leq\sigma_1,
\tag{P1}
\end{equation}
\end{minipage}%
\begin{minipage}{.5\linewidth}
\begin{equation}\label{P2}
\PX[L(q)>C(q)]\leq\sigma_2.
\tag{P2}
\end{equation}
\end{minipage}
\par\vspace{5pt}
\noindent \eqref{P1} is solvency, the stricter tail event, and \eqref{P2} is \ac{LP} non-loss, requiring premiums instead of reserve to cover ordinary payouts. Neither bounds severity, and a shared reserve is exhausted by rare severe tails, so we strengthen both to \ac{CVaR} constraints.

\textit{Least-premium cost function.} We strengthen \eqref{P2} first, requiring the premium to cover the average loss over the worst $\sigma_2$-fraction of outcomes. For a nonnegative payout $Y$, this tail average is the upper-tail conditional value-at-risk \cite{rockafellar2000optimization,nemirovski2006convex},
\[
\CVaR_\sigma^{\PX}(Y):=\min_{\eta\in\R}\Big\{\eta+\tfrac{1}{\sigma}\E_{\PX}[(Y-\eta)_+]\Big\},
\]
where the associated upper-tail quantile is $\VaR_\sigma^{\PX}(Y):=\inf\{a:\PX[Y\leq a]\geq 1-\sigma\}$.
Since a tail average is at least its threshold, a \ac{CVaR} bound implies the chance constraint \[\CVaR_\sigma^{\PX}(Y)\leq a \quad\Longrightarrow\quad \PX[Y>a]\leq\sigma .\]
The \ac{CVaR} bound is also convex in $q$. The \ac{LP} non-loss surrogate is \eqref{P2'}, which implies \eqref{P2}, so any such $C$ clears the pointwise floor $C(q)\geq \CVaR_{\sigma_2}^{\PX}(q\cdot X)$. Full-price selling, $C(q)=\mathbf{1}^\top q$, is feasible but accepts unbounded notional, making capacity maximization ill-posed, so we impose a least-premium principle:
\par\vspace{5pt}\noindent
\begin{minipage}{.5\linewidth}
\begin{equation}\label{P2'}
\CVaR_{\sigma_2}^{\PX}(L(q))\leq C(q),
\tag{P2'}
\end{equation}
\end{minipage}%
\begin{minipage}{.5\linewidth}
\begin{equation}
C^*(q):=\CVaR_{\sigma_2}^{\PX}(q\cdot X).
\label{eq:puamm-cost}
\end{equation}
\end{minipage}
\par\vspace{5pt}

This floor is usable because \ac{CVaR} is convex and positively homogeneous in $q$ (\S~\ref{sec:basic-properties}), whereas the $\VaR_{\sigma_2}$ floor under raw \eqref{P2} is nonconvex in general. The quote $C^*(q+\delta e_i)-C^*(q)$ is a portfolio underwriting price, not a belief price: for a single long shot the empty-book unit quote is $\min(1,p_i/\sigma_2)$${}>p_i$, reflecting undiversified tail risk, and on diversified books the loading shrinks toward fair value (\S~\ref{sec:capital-efficiency}).

\subsection{Deriving the Acceptance Rule}

Solvency then decides acceptance, and
the same implication strengthens \eqref{P1} to
\begin{equation}
\CVaR_{\sigma_1}^{\PX}(L(q))\leq C(q)+K ,
\tag{P1'}\label{P1'}
\end{equation}
which implies \eqref{P1}. With $C=C^*$, the residual object is the \emph{tail-gap} function
\begin{equation}
f(q):=
\CVaR_{\sigma_1}^{\PX}(q\cdot X)
-
\CVaR_{\sigma_2}^{\PX}(q\cdot X),
\label{eq:tail-gap}
\end{equation}
which is nonnegative because $\sigma_1<\sigma_2$. Under $C^*$, \eqref{P1'} holds exactly when
$f(q)\leq K$, so the acceptance rule is
\begin{equation}
\tau^*(q):=\mathbf{1}\{f(q)\leq K\},
\qquad
\mathcal{Q}(K):=\{q\in\R_{\geq 0}^n:f(q)\leq K\}.
\label{eq:puamm-acceptance}
\end{equation}

\begin{mdframed}[style=pumbox]
\begin{definition}[Portfolio underwriting AMM]
\label{def:puamm}
Given a joint distribution $\PX$, reserve capital $K$, and tail levels $0<\sigma_1<\sigma_2<1$, the \textup{\name} mechanism is the cost-function \ac{AMM} $(C^*,\tau^*)$ defined by \eqref{eq:puamm-cost}, \eqref{eq:tail-gap}, \eqref{eq:puamm-acceptance}. A requested trade that moves the state to $q'$ pays $C^*(q')-C^*(q)$ and is accepted exactly when $q'\in\mathcal{Q}(K)$.
\end{definition}
\end{mdframed}

The tail gap is the risk that premiums leave unfunded, so the reserve backs only the $\sigma_1$--$\sigma_2$ increment. Among approximations of a chance constraint that imply it and are built from a one-dimensional generating function, the \ac{CVaR} bound is the tightest \cite{nemirovski2006convex}, so \eqref{P1'} and \eqref{P2'} are the least conservative such surrogates for \eqref{P1} and \eqref{P2}.

\subsection{The Price--Capacity Frontier}

The derived pair is not one feasible design but the boundary of
the feasible set. Call $(C,\tau)$ \emph{feasible} if $C$ is valid (\S~\ref{sec:model}) and accepted states satisfy \eqref{P1'} and \eqref{P2'}.

\begin{theorem}[Feasibility frontier]
\label{thm:frontier}
Let $(C,\tau)$ be feasible. \textup{(a)} Every accepted state $q$ satisfies
$C(q)\geq C^*(q)$. \textup{(b)} If $C=C^*$, the maximal acceptance rule
consistent with \eqref{P1'} is $\tau^*$. \textup{(c)} If $(C,\tau)$
accepts some state $q$ with $f(q)>K$, then \[C(q)-C^*(q)\;\geq\;f(q)-K\;>\;0 .\]
\end{theorem}

\noindent\textit{Proof.} (a) is \eqref{P2'} at $q$ and the definition of $C^*$. (b) Under $C^*$, \eqref{P1'} holds at $q$ iff $f(q)\leq K$. (c) By \eqref{P1'}, $C(q)\geq\CVaR_{\sigma_1}^{\PX}(q\cdot X)-K=C^*(q)+f(q)-K$.

Premiums cannot fall below $C^*$ at any accepted state (a floor on collected totals, not on each marginal quote), and acceptance cannot extend beyond $\mathcal{Q}(K)$ except by overcharging at least the excess tail gap. Full-price selling sits on the overcharge branch and \name{} on the refusal branch, so least premium is the trader-favoring endpoint rather than an extra normative choice. Parts \textup{(a)} and \textup{(c)} use only \eqref{P1'} and \eqref{P2'} at accepted states, not the validity axioms, so the frontier is not an artifact of the cost-function class. Under bare \eqref{P2} the floor is only $\VaR_{\sigma_2}^{\PX}(q\cdot X)$, and valid schedules may undercut $C^*$ while controlling only \ac{LP}-loss frequency.

\subsection{Basic Properties of \name}
\label{sec:basic-properties}

\begin{proposition}
\label{prop:validity}
The cost function $C^*$ is convex, path-independent, satisfies $C^*(0)=0$,
and the price bounds \eqref{eq:valid-price}. Every
accepted state $q\in\mathcal{Q}(K)$ satisfies \eqref{P2'} and \eqref{P1'} with $C=C^*$,
hence the chance constraints \eqref{P2} and \eqref{P1}.
\end{proposition}

\noindent\textit{Proof (sketch).} Convexity and path independence are immediate. The price bounds follow from $L\leq L+\delta X_i\leq L+\delta$ by monotonicity and translation invariance of \ac{CVaR}. \eqref{P2'} holds with equality, and on $\mathcal{Q}(K)$ adding $f(q)\leq K$ gives \eqref{P1'} (Appendix~\ref{app:proofs}).

\textit{Prices.} By finite linear-programming duality, $C^*$ and $f$ are maxima of $\E_Q[q\cdot X]$ over the reweighted tail sets $\mathcal{D}_\sigma:=\{Q:\ Q(\omega)\leq\PX(\omega)/\sigma\ \forall\omega\}$ at levels $\sigma_2$ and $\sigma_1$ (Proposition~\ref{prop:dual}, Appendix~\ref{app:proofs}). Marginal prices, and subgradients in general, are therefore expectations $\E_Q[X]$ under adversarially reweighted tails and lie in $[0,1]^n$.

\textit{Axiom audit.}
\label{sec:audit}
\name{} keeps path independence, the price bounds \eqref{eq:valid-price}, and demand responsiveness. It gives up always-quoting and deterministic solvency, the source of leverage, and its induced scoring rule is proper but not strictly proper, so \name{} is a tail-liquidity backstop rather than a price-discovery venue by theorem (Appendix~\ref{app:audit}).

\textit{Proportional-haircut settlement.}
When $L(q)>C^*(q)+K$, the \ac{AMM} settles by a proportional haircut: each winning claim receives
$h(X):=(C^*(q)+K)/L(q)\in(0,1)$ of its unit payout, so aggregate settlement is
$C^*(q)+K$. The haircut settles the deep tail that the reserve does not cover, with probability at most $\sigma_1$ on accepted states.

\section{Capital Efficiency Analysis}
\label{sec:capital-efficiency}

This section quantifies leverage, the YES notional $\mathbf{1}^\top q$ that one unit of reserve can back, which is bounded by the tail gap $f(q)$ the reserve must cover.

\subsection{Ratio Form}

Define the capital utilization of \name at reserve level $K>0$ by
\begin{equation}
U^*(K):=
\frac{1}{K}
\sup_{q\in\mathcal{Q}(K)}
\mathbf{1}^\top q.
\label{eq:capital-utilization}
\end{equation}
Thus $U^*(K)$ is the maximum accepted YES notional per reserve unit. Let
$\Delta_n:=\{u\in\R_{\geq 0}^n:\mathbf{1}^\top u=1\}$ be the probability
simplex, and define
\begin{equation}
m(\PX):=\min_{u\in\Delta_n}
\left[
\CVaR_{\sigma_1}^{\PX}(u\cdot X)
-
\CVaR_{\sigma_2}^{\PX}(u\cdot X)
\right].
\label{eq:tail-gap-minimum}
\end{equation}

\begin{theorem}[Capital-utilization ratio form]
\label{thm:ratio-form}
If $m(\PX)>0$, then
\[
U^*(K)
=
\sup_{q\geq 0,\ q\neq 0}
\frac{\mathbf{1}^\top q}{f(q)}
=
\frac{1}{m(\PX)}.
\]
In particular, $U^*(K)$ is independent of $K$. If $m(\PX)=0$,
utilization is unbounded along a zero-gap direction where premiums
alone carry the stricter tail, a degenerate case.
\end{theorem}

\noindent\textit{Proof (sketch).} By positive homogeneity, writing $q=(\mathbf{1}^\top q)u$ with $u\in\Delta_n$, the constraint reads $\mathbf{1}^\top q\leq K/f(u)$, so $U^*(K)=$ $1/\min_{\Delta_n}f=1/m(\PX)$.

Problem \eqref{eq:tail-gap-minimum}, a difference of two \ac{CVaR} functions, is generally nonconvex.

\begin{corollary}[Service capacity]
\label{cor:capacity}
Assume $m(\PX)=\min_{u\in\Delta_n}f(u)>0$. Along any sequence of accepted trades
from $q=0$, every intermediate state satisfies
$\mathbf{1}^\top q\leq U^*K$. Conversely,
for any $\varepsilon\in(0,1)$, single-market trades of size at most
$\varepsilon K$ can reach $(1-\varepsilon)U^*K$, and a bundle trade in an
optimal direction attains $U^*K$ exactly. Thus $U^*K$ is the order-free ceiling
on the tail-side notional the \ac{AMM} can serve.
\end{corollary}

\noindent\textit{Proof (sketch).} The upper bound combines $f(q)\geq(\mathbf{1}^\top q)\,m$ with acceptance of every prefix; tightness uses a round-robin construction along an optimal direction and the increment bound $f(q+\delta e_i)\leq f(q)+\delta$. The complete proof is in Appendix~\ref{app:proofs}.

\subsection{An i.i.d. Benchmark}

The ratio form makes diversification explicit. Let $X_1,\ldots,X_n$ be i.i.d.\
Bernoulli$(p)$ in the regime $\sigma_2>p$: a single market has
$\CVaR_\sigma(X_i)=\min(1,p/\sigma)$, and if $\sigma_2\leq p$ then $f(e_i)=0$,
the degenerate case of Theorem~\ref{thm:ratio-form}. Write
$S_n=\sum_{i=1}^n X_i$ and $u^{\mathrm{sym}}=\mathbf{1}/n$.

\begin{corollary}[Symmetric i.i.d. allocation]
\label{cor:iid-symmetric}
Under the i.i.d. Bernoulli benchmark with $\sigma_2>p$, the symmetric
direction achieves the following utilization:
\begin{equation}
U_{\mathrm{sym}}(n,p,\sigma_1,\sigma_2)
=
\frac{n}{
\CVaR_{\sigma_1}(S_n)-\CVaR_{\sigma_2}(S_n)
}.
\label{eq:symmetric-utilization}
\end{equation}
Consequently, $U^*(K)\geq U_{\mathrm{sym}}(n,p,\sigma_1,\sigma_2)$. As
$n\to\infty$ with $p$ fixed,
\begin{equation}
U_{\mathrm{sym}}(n,p,\sigma_1,\sigma_2)
=
\frac{\sqrt n}{
\sqrt{p(1-p)}\,c(\sigma_1,\sigma_2)
}
\cdot(1+o(1)),
\label{eq:symmetric-sqrt-n}
\end{equation}
where, with $\varphi,\Phi$ the standard normal pdf and cdf, the Gaussian constants are
\[
c(\sigma_1,\sigma_2)
:=
\frac{\varphi(z_{\sigma_1})}{\sigma_1}
-
\frac{\varphi(z_{\sigma_2})}{\sigma_2},
\qquad
z_\sigma:=\Phi^{-1}(1-\sigma).
\]

\end{corollary}

\looseness=-1
\noindent\textit{Proof (sketch).} $u^{\mathrm{sym}}\cdot X=S_n/n$, so homogeneity gives $f(u^{\mathrm{sym}})=\frac1n[\CVaR_{\sigma_1}(S_n)-\CVaR_{\sigma_2}(S_n)]$. The asymptotics are the central limit theorem with uniform integrability, and subtracting the two levels cancels the mean term (Appendix~\ref{app:proofs}).

We state this as a lower bound because the symmetric direction is not always optimal at finite $n$: at $(n,p,\sigma_1,\sigma_2)=(3,0.05,0.01,0.10)$ the direction $(0,\tfrac12,\tfrac12)$ certifies $U^*\geq8$ against $U_{\mathrm{sym}}\approx4.52$, since the binomial tail is a staircase and concentrating exposure can place both \ac{CVaR} levels on one atom. In the exact enumeration of Appendix~\ref{app:numerics} ($n\leq12$), the advantage peaks at $n=4$ ($81\%$) and vanishes at $n=12$.

Corollary~\ref{cor:iid-symmetric} thus certifies $U^*=\Omega(\sqrt n)$, grows at a minimum speed of $\sqrt n$. In this paper every headline capacity is the constructive certificate $U_{\mathrm{sym}}$. Whether $U^*$ grows faster than $\sqrt n$ remains open, and a faster rate would only favor us and add capacity.

This leverage is not bought by overcharging: along the symmetric ray the
\emph{average unit price} $\tfrac1n\CVaR_{\sigma_2}(S_n)
=p+\sqrt{p(1-p)/n}\,\varphi(z_{\sigma_2})/\sigma_2+o(n^{-1/2})$ tends to fair
value $p$, so $\Omega(\sqrt n)$ leverage comes with a vanishing
$\Theta(n^{-1/2})$ markup, unlike never-default designs, which need unit prices near one for comparable leverage
(Proposition~\ref{prop:barrier}).

\subsection{Robustness to TV-Small Dependence}

The i.i.d.\ benchmark is a reference point. Utilization is stable under
departures from the independent product that are small in total variation
$\TV(\mathbb P,\mathbb Q):=\sup_A|\mathbb P(A)-\mathbb Q(A)|$. The workhorse
is a perturbation lemma: for a loss in $[0,M]$, \ac{CVaR} at level $\sigma$
moves by at most $(M/\sigma)\,\TV(\mathbb P,\mathbb Q)$
(Lemma~\ref{lem:cvar-tv}, Appendix~\ref{app:proofs}).

Let $\PX^0=\prod_i\mathbb{P}_{X_i}$ be the independent product with the same
marginals as $\PX$, and set $\epsilon:=\TV(\PX,\PX^0)$. For a law $\mathbb{P}$
on $\{0,1\}^n$, write $f^{\mathbb{P}}(u)$ for the tail gap of direction $u$ and
$m(\mathbb{P}):=\min_{u\in\Delta_n}f^{\mathbb{P}}(u)$, both computed under
$\mathbb{P}$, and, when $m(\mathbb{P})>0$,
$U^*(\mathbb{P}):=1/m(\mathbb{P})$ for the ratio-form utilization of
Theorem~\ref{thm:ratio-form}.

\begin{theorem}[TV robustness of utilization]
\label{thm:tv-robustness}
Let $m_0:=m(\PX^0)>0$ and
$\alpha:=\epsilon(1/\sigma_1+1/\sigma_2)$. Then
$|m(\PX)-m_0|\leq \alpha$. If $\alpha<m_0$, then
\[
\frac{U^*(\PX^0)}{1+\alpha/m_0}
\leq
U^*(\PX)
\leq
\frac{U^*(\PX^0)}{1-\alpha/m_0}.
\]
In particular, if $\alpha\leq m_0/2$, then
$\bigl|U^*(\PX)/U^*(\PX^0)-1\bigr|
\leq 2\alpha/m_0
\leq 4\epsilon/(\sigma_1 m_0)$.
\end{theorem}

The proof applies the perturbation lemma at both tail levels, uniformly over the simplex, and inverts via Theorem~\ref{thm:ratio-form} (Appendix~\ref{app:proofs}). The bound imposes no structure on the dependence, and in the common-shock experiments of Appendix~\ref{app:numerics} the realized shift is $10.8$ to $24.8$ times smaller than the bound, because both \ac{CVaR} levels move together. Corollary~\ref{cor:default-robust} needs neither a smallness condition nor any structure on the error.

\begin{corollary}[Default-probability robustness]
\label{cor:default-robust}
Let $\mathbb{Q}$ be the true law and $\epsilon_{\mathbb{Q}}:=\TV(\PX,\mathbb{Q})$. At every state $q$ accepted under $\PX$, $\mathbb{Q}[L(q)>C^*(q)+K]\leq\sigma_1+\epsilon_{\mathbb{Q}}$ and $\mathbb{Q}[L(q)>C^*(q)]\leq\sigma_2+\epsilon_{\mathbb{Q}}$. Proof in Appendix~\ref{app:proofs}.
\end{corollary}

\section{Deployment in Rare-Event Prediction Markets}
\label{sec:application}

\name{} runs alongside a primary venue, an \ac{LMSR}-style \ac{AMM}, pm-AMM, or a \acl{CLOB} (\acs{CLOB}, e.g., Polymarket)\acused{CLOB}, which supplies prices, and serves the block-sized tail demand the primary book fills only at a steep markup.

\subsection{Trader Routing}

\paragraph{Primary execution price.} Page and Clemen \cite{page2013calibrated} show on over $500{,}000$ transactions that long-horizon prices exceed calibrated probabilities on the rare side, which their model attributes to the cost of capital committed until resolution. We therefore model the primary unit price as $\pi^P_i=p_i+\kappa_i\in[0,1]$ with $\kappa_i\geq0$, and the \emph{carry-only benchmark} takes $\kappa_i=rT_i$ for effective horizon $T_i$ and per-unit-time capital cost $r$. For a block trade, the relevant price is the average price of the primary liquidity the block would consume, which includes carry, inventory, adverse-selection, fee, and book-walking costs.

Let $P_i(s)$ be the minimum cost of buying $s\geq0$ YES claims in market $i$ at the primary venue, with $P_i(0)=0$ and $P_i(s)=+\infty$ if unavailable. A trader with parent order $\delta$ sends a feasible tranche $x\in\mathcal{A}_i(q,\delta):=\{x\in[0,\delta]:q+xe_i\in\mathcal{Q}(K)\}$ to \name{} and $\delta-x$ to the primary venue, at total cost $\mathcal{E}_i(q,\delta,x):=C^*(q+xe_i)-C^*(q)+P_i(\delta-x)$. Routing a tranche $x>0$ weakly improves on all-primary execution exactly when the average tranche price does not exceed the average price of the primary block it replaces,
\begin{equation}
\pi_{\mathrm{PU},i}(q,x):=\frac{C^*(q+xe_i)-C^*(q)}{x}
\;\leq\;
\bar\pi^P_i(\delta,x):=\frac{P_i(\delta)-P_i(\delta-x)}{x},
\label{eq:block-routing}
\end{equation}
or, subtracting $p_i$ from both sides, $\pi_{\mathrm{PU},i}(q,x)-p_i\leq\bar\kappa^P_i(\delta,x):=\bar\pi^P_i(\delta,x)-p_i$. In a \ac{CLOB}, $\bar\pi^P_i(\delta,x)$ is the average price of the last $x$ units an all-primary order would consume, which exceeds the best ask whenever the order walks the book. Proposition~\ref{prop:block-cutoff} locates the cutoff on standard primary curves and shows how it moves with the book.

\begin{proposition}[Block-premium cutoffs and self-limiting advantage]
\label{prop:block-cutoff}
Fix $q\in\mathcal{Q}(K)$, a market $i$, an order $\delta$ with
$P_i(\delta)<\infty$, and a feasible tranche $x\in\mathcal{A}_i(q,\delta)$, $x>0$.
\textup{(a)} If the primary execution curve is linear, $P_i(s)=(p_i+\kappa_i)s$, routing weakly improves exactly when
$\kappa_i\geq\pi_{\mathrm{PU},i}(q,x)-p_i$. Under the carry-only
benchmark $\kappa_i=rT_i$ with $r>0$, this reads
$T_i\geq(\pi_{\mathrm{PU},i}(q,x)-p_i)_+/r$.
\textup{(b)} At the empty \name{} book, for
$0<p_i\leq\sigma_2$, every accepted $x>0$ has tranche price $\pi_{\mathrm{PU},i}(0,x)=p_i/\sigma_2$, so routing improves exactly when $\bar\kappa^P_i(\delta,x)\geq p_i(1-\sigma_2)/\sigma_2$.
\textup{(c)}
$\pi_{\mathrm{PU},i}(q,x)$ is nondecreasing in $x$ and in the base point along
$e_i$, so the \name{} execution advantage is self-limiting. Proof in Appendix~\ref{app:proofs}.
\end{proposition}

The cutoff compares execution prices only. A \name{} tranche also carries the haircut backstop of \S~\ref{sec:basic-properties}, which binds with probability at most $\sigma_1$, so a risk-neutral trader's expected haircut loss is at most $\sigma_1$ per unit (Proposition~\ref{prop:validity}) and the cutoff \eqref{eq:block-routing} shifts by at most $\sigma_1$: one cent per unit at $(\sigma_1,\sigma_2)=(0.01,0.10)$ and $p_i=0.02$, against an empty-book cutoff of $p_i(1-\sigma_2)/\sigma_2=0.18$. \name{} therefore wins where primary order books are expensive or thin, as \S~\ref{sec:numerics} confirms on Polymarket.

\subsection{LP Expected Profit}

The \ac{LP}'s terminal profit is
$\Pi_L(q,X):=C^*(q)-q\cdot X$. Haircuts only raise the realized profit, to $C^*(q)-\min(q\cdot X,\,C^*(q)+K)$, so the formulas below are lower bounds.

\begin{proposition}[Conditional premium cushion]
\label{prop:lp-apr}
For every \name-accepted state $q$,
\[
\E_{\PX}[\Pi_L(q,X)]
=
\CVaR_{\sigma_2}^{\PX}(q\cdot X)-\E_{\PX}[q\cdot X]
\geq0,
\]
with strict inequality whenever $q\cdot X$ is non-degenerate. For a direction
$u\in\Delta_n$ with $f(u)>0$, let
$a(u):=\CVaR_{\sigma_2}^{\PX}(u\cdot X)-\E_{\PX}[u\cdot X]$ be its premium cushion. The state $q_{\phi,u}:=\phi K u/f(u)$, which uses a fraction $\phi\in[0,1]$ of the ray capacity $K/f(u)$, satisfies
\[
\E_{\PX}[\Pi_L(q_{\phi,u},X)]
=
\phi K\,\frac{a(u)}{f(u)}.
\]
\end{proposition}

Appendix~\ref{app:proofs} gives the proof and the induced LP APR over an exogenous horizon.

\subsection{Inputs, Implementation, and Attacks}
\label{sec:threat}

\textit{Sourcing the input.} The marginals $p_i$ can come from several sources. The simplest is the primary venue's mid-price, taken as a \ac{TWAP} so that a transient distortion moves $p_i$ only in proportion to its duration. The ask would overstate $p_i$, since it includes carry, inventory, and adverse-selection loadings, and long-shot prices exceed realized frequencies even for buyers posting their own bids \cite{cardozo2026longshot}, as do prices in our data (\S~\ref{sec:numerics}), which makes the input conservative for the reserve. Other sources are professional market makers' models~\cite{feil2026optimalmarketmakingprediction}, bookmaker odds~\cite{strumbelj2014determining}, aggregated forecasting polls \cite{atanasov2017distilling}, and, for events defined by an asset price, option-implied (risk-neutral) probabilities \cite{breeden1978prices}. Dependence comes from historical co-movement, a factor model, or a learned joint model \cite{rana2026parlaymarket}, and capacity depends on its upper tail, the probability that many listed markets resolve YES together. These choices leave $C^*$ and $\tau^*$ unchanged, and input error moves certified capacity only as in Theorem~\ref{thm:tv-robustness}.

\textit{On-chain and off-chain computation.} The contract stores only the state $q$ and the reserve $K$. Because $C^*$ and $f$ depend on $q$ only through the scalar law of $q\cdot X$, an oracle can compute them off chain and post each quote with its acceptance bit. Both are linear programs under a scenario representation \cite{rockafellar2000optimization}, our inputs admit exact convolution, and approximate oracles need a safety margin near the boundary $f(q)=K$ (Appendix~\ref{app:numerics}).

\textit{Trust model.} Three parties are trusted: the operator that supplies $\PX$, the oracle that evaluates $C^*$ and $f$, and the resolution source (same as the primary venue). The oracle is checkable: with $\PX$, anyone can recompute the quote and acceptance. An error of total variation $\epsilon$ in $\PX$ raises the default probability by at most $\epsilon$ (Corollary~\ref{cor:default-robust}). \acp{LP} can audit the published law before committing, as lending-protocol users vet a price oracle.

\textit{Adverse selection.} Every liquidity provider faces informed flow: dealers price it into their spreads \cite{glosten1985bid}, and an \ac{LMSR}'s bounded loss is the subsidy it pays to better-informed traders~\cite{hanson2007lmsr}. \name{} prices from $\PX$ and does not learn from order flow, so a trader who knows more than $\PX$ buys the claims it underprices. Its \ac{CVaR} loading acts like a spread, and convex quotes plus the cap $f(q)\leq K$ bound what such a trader can take. Solvency is robust to such flow: Corollary~\ref{cor:default-robust} holds at every accepted state, whoever chose it, so the true default probability stays below $\sigma_1+\TV(\PX,\mathbb{Q})$.
A defense is to make the input accurate, including a refreshed primary venue price or a more professional source.

\textit{Input manipulation.} When marginals come from primary mid-prices, moving a thin primary mid moves \name{} quotes. A \ac{TWAP} mid and a per-market halt on fast mid moves defend against this, and both only refuse trades, so every statewise guarantee survives. Manipulation cannot force settlement, since claims settle only at resolution. Strategic timing of trades, such as front-running~\cite{chan2025mechanism,daian2020flash}, is outside the model.

\subsection{Listing and Cold Start}
\label{sec:listing}

Which markets to list is an operator decision, but the mechanism scores it exactly: a candidate book is certified at $K/m(\PX)$ (Theorem~\ref{thm:ratio-form}), and listing a market changes no existing quote and never lowers certified capacity (Lemma~\ref{lem:listing}, Appendix~\ref{app:selection}). A greedy procedure follows: given a candidate set and the operator's joint law, repeatedly add the market whose inclusion raises $K/m$ the most, or, as a cheaper proxy, raises the tail gap along the symmetric ray the least. Since no candidate lowers capacity, the ranking is decided by the joint upper tail the operator supplies, and what counts is the number of independent clusters rather than the number of markets: two independent comonotone blocks of $50$ markets certify $U=8.00$ against $39.26$ for $100$ i.i.d.\ markets at $p=0.05$ (Table~\ref{tab:app-examples}). Balancing the book across causally unrelated categories is a practical proxy, and clusters could be detected from platform tags or a classifier. Arbitrageurs can balance the prices between \name{} and the primary venue.

The operator also chooses the tail levels: a larger $\sigma_1$ buys capacity at a higher default probability, and a larger $\sigma_2$ lowers premiums and capacity and raises the \ac{LP}-loss bound. The empty-book quote is $p_i/\sigma_2$ (Proposition~\ref{prop:block-cutoff}(b)), so small orders stay primary until \name{} diversifies, as any symmetric seed achieves (Appendix~\ref{app:bootstrap}).

\section{Evaluation}
\label{sec:numerics}

Section~\ref{sec:numerical} tests \name{} across market counts and dependence and compares it with capital-matched \acp{LMSR}, the canonical prediction market \ac{AMM}, and \S~\ref{sec:empirical} compares it with the Polymarket \ac{CLOB} and replays it on $30$ months of resolutions. By default, $(\sigma_1,\sigma_2)=(0.01,0.10)$, $K=1$, and all CVaRs are exact finite-law calculations.

\subsection{Numerical Evaluation}
\label{sec:numerical}

\paragraph{Capacity and prices.}
Table~\ref{tab:num-capacity-price} reports the symmetric capacity certificate
under i.i.d.\ Bernoulli markets. At $(n,p)=(100,0.02)$, one reserve unit backs $55.34$ units of tail-side notional against $1.02$ under full collateral, and the average price at the boundary is $2.36$ times fair value. Across the table, capacity grows with $n$ while this markup falls, as Corollary~\ref{cor:iid-symmetric} predicts. At every boundary state, the supplied-law insolvency probability is $0.0014$ to $0.0072$ and the \ac{LP}-loss probability $0.013$ to $0.051$, inside $\sigma_1$ and $\sigma_2$. On the insolvency event the haircut returns $78\%$ to $97\%$ of the unit payout (Appendix~\ref{app:numerics}).

\begin{table}[t]
\centering
\caption{Independent-market capacity and boundary prices. Ratio is $U_{\mathrm{sym}}$ over the full-collateral ceiling $1/(1-p)$, and price$/p$ is the average unit price at the boundary over fair value.}
\label{tab:num-capacity-price}
\small
\setlength{\tabcolsep}{3.1pt}
\begin{tabular}{c@{\quad}ccc@{\quad}ccc@{\quad}ccc}
\toprule
\multicolumn{1}{c}{} & \multicolumn{3}{c}{$p=0.02$}
& \multicolumn{3}{c}{$p=0.05$}
& \multicolumn{3}{c}{$p=0.10$}\\
\cmidrule(lr){2-4}\cmidrule(lr){5-7}\cmidrule(lr){8-10}
$n$ & $U_{\mathrm{sym}}$ & ratio & price$/p$
& $U_{\mathrm{sym}}$ & ratio & price$/p$
& $U_{\mathrm{sym}}$ & ratio & price$/p$\\
\midrule
$25$  & $22.26$  & $21.8\times$  & $4.07$
      & $17.48$  & $16.6\times$  & $2.74$
      & $13.95$  & $12.6\times$  & $2.17$\\
$100$ & $55.34$  & $54.2\times$  & $2.36$
      & $39.26$  & $37.3\times$  & $1.82$
      & $30.53$  & $27.5\times$  & $1.55$\\
$400$ & $129.61$ & $127.0\times$ & $1.66$
      & $88.74$  & $84.3\times$  & $1.40$
      & $66.79$  & $60.1\times$  & $1.27$\\
\bottomrule
\end{tabular}
\end{table}

\begin{figure}[t]
\centering
\includegraphics[width=\textwidth]{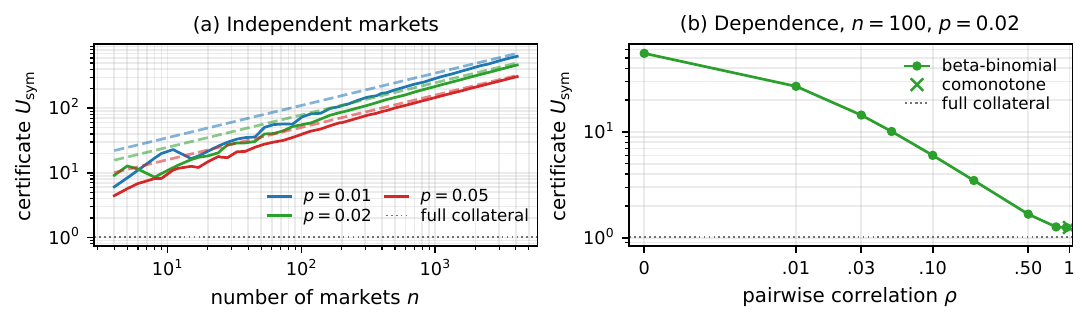}
\caption{Capacity experiments. (a) Exact independent-market scaling, Gaussian $\sqrt n$ asymptote dashed. (b) Beta-binomial dependence at $(n,p)=(100,0.02)$, comonotone endpoint at $\rho=1$.}
\label{fig:num-capacity-dependence}
\end{figure}

Figure~\ref{fig:num-capacity-dependence}(a) shows staircases around $\sqrt n$ scaling, and (b) replaces independence by an exchangeable beta-binomial law with pairwise
correlation $\rho$. At $(n,p)=(100,0.02)$, the certificate falls from $55.34$ at $\rho=0$
to $27.00$ at $\rho=0.01$, $6.01$ at $\rho=0.10$, and $1.67$ by $\rho=0.50$,
near the comonotone endpoint $1.25$ \cite{dhaene2002comonotonicity}.

\begin{figure}[t]
\centering
\begin{minipage}[t]{0.485\textwidth}
\centering
\includegraphics[width=\linewidth]{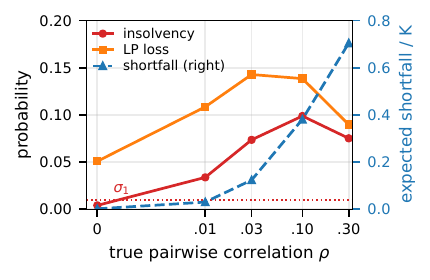}
\caption{Misspecified dependence. A book certified as independent at $(n,p)=(100,0.02)$ under true beta-binomial correlation $\rho$: frequencies (left axis) and expected shortfall (right).}
\label{fig:misspec}
\end{minipage}\hfill
\begin{minipage}[t]{0.485\textwidth}
\centering
\includegraphics[width=\linewidth]{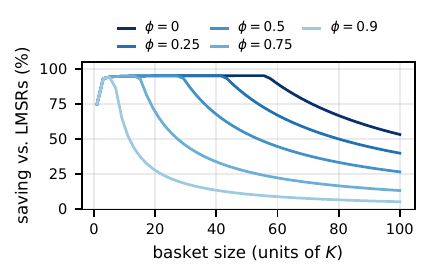}
\caption{Capital-matched \ac{AMM} baseline ($n=100$, $p=0.02$): basket saving over \acp{LMSR} with total worst-case loss $K$, bending at the remaining capacity $(1-\phi)U_{\mathrm{sym}}K$.}
\label{fig:lmsr}
\end{minipage}
\end{figure}

\paragraph{Misspecified dependence.}
Misspecification bites in one direction only. We certify the book on the symmetric ray at $(n,p)=(100,0.02)$ under a supplied correlation $\rho_s$, place it at the boundary, and evaluate it under a true correlation $\rho$ (Figure~\ref{fig:misspec}). Understating dependence is costly: a book certified as independent ($0.0041$ insolvency probability under the supplied law) reaches $0.034$ at $\rho=0.01$, since with $100$ markets a correlation of $0.01$ already doubles the variance of the YES count. Overstating is safe and cheap: for $\rho_s$ up to $0.10$, every true $\rho\leq\rho_s$ keeps insolvency probability below $\sigma_1$, and a book certified at $\rho_s=0.01$ still backs $27.00$ units per reserve unit, $26$ times full collateral (Table~\ref{tab:app-misspec}, Appendix~\ref{app:numerics}), so conservative inputs trade capacity for solvency. From $\rho=0.10$ to $0.30$, stronger dependence moves risk from frequency to severity: insolvency probability falls while expected excess loss rises, because positive correlation pushes the aggregate loss toward both extremes. Resolved Polymarket long shots show no such clustering (\S~\ref{sec:empirical}).

\textit{LP cushion at the boundary.}
At the symmetric boundary for $(n,p)=(100,0.02)$, Proposition~\ref{prop:lp-apr} gives premiums $2.61K$ on notional $55.34K$, expected payout $1.11K$, and expected \ac{LP} profit $1.50K$, with $\Pr[W<K]=0.0508$ and $\Pr[W<0]=0.0041$.

\textit{Capital-matched AMM baseline.} When the primary venue is an \ac{AMM}, the comparison is exact: against $100$ independent \acp{LMSR} whose subsidies sum to the same reserve (each $b=(K/n)/\log(1/p)$, total worst-case loss $K$), routing a basket of $50K$ contracts through \name{} at half utilization costs $53.8\%$ less, since each \ac{LMSR}'s depth is fixed by its share of the reserve while \name{} pools it (Figure~\ref{fig:lmsr}). Section~\ref{sec:empirical} runs the same comparison against real \ac{CLOB} depth.

\subsection{Empirical Evaluation}
\label{sec:empirical}

\begin{figure}[t]
\centering
\includegraphics[width=\textwidth]{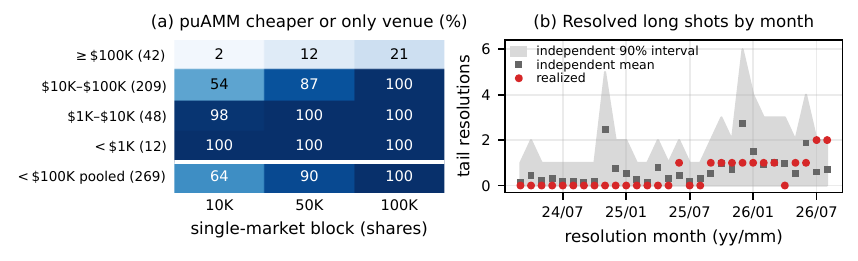}
\caption{Polymarket evidence. (a) Share of markets in which a \name{} tranche ($K=\$100\mathrm{K}$, half-utilized $311$-market book) is cheaper than the ladder or the only venue, by resting liquidity. (b) Monthly tail resolutions of $907$ long shots against the independent $90\%$ interval.}
\label{fig:empirical}
\end{figure}

\paragraph{Polymarket order books.} We snapshot the order books of every active Polymarket market whose long-shot side traded between $0.5$ and $5$ cents with at least two weeks to resolution, $673$ of $11{,}038$ active markets on 2026-09-24 (data rules in Appendix~\ref{app:numerics}). By median execution price relative to the mid (Figure~\ref{fig:app-depth}), the $159$ markets with at least \$100K of resting liquidity are deep: 10K shares cost $1.12\times$ the mid. The other $514$ are thin: 10K shares cost $6.6\times$ the mid at the median, $27$ cannot fill that order at all, and none fills 100K shares below twice the mid. Every share of that depth is fully collateralized. We then list one market per event ($311$ markets, $\sum_i p_i\approx8$) with $p_i$ at the mid and independent $\PX$, place the \name{} book at half capacity with $K=\$100\mathrm{K}$, and price a single-market block of $x$ shares on both venues. We do not test dependence among these events, which a deployed book would account for when selecting markets (\S~\ref{sec:listing}). Outside the most liquid quartile, \name{} is cheaper than walking the primary order book, or is the only venue able to fill, in $64\%$ of markets at 10K shares, $90\%$ at 50K, and $100\%$ at 100K (Figure~\ref{fig:empirical}(a)). This is the pattern Proposition~\ref{prop:block-cutoff} predicts: the symmetric-ray premium is $1.63\times$ the mean mid, so \name{} loses at the touch and wins where the book runs out. A diversified basket over the book saves $18\%$ at 10K shares and $27\%$ at 50K against all-primary execution, and from 100K shares the primary book cannot fill it. Ten snapshots agree: $92\%$ to $93\%$ of long shots cannot fill 100K shares below twice the mid, and outside the most liquid quartile \name{} is cheaper or the only venue in $64\%$ to $76\%$ of markets at 10K shares and at least $98\%$ at 100K.

\paragraph{Resolved long shots.} Resolutions test whether independence understates how often rare outcomes coincide. We take the $907$ Polymarket markets resolved between January $2024$ and August $2026$ whose long-shot side had a seven-day \ac{TWAP} $p_i\in[0.005,0.05]$ thirty days before resolution (volume, lifetime, one-per-event, and month-size rules in Appendix~\ref{app:numerics}), and group them by resolution month ($30$ months). Prices do not understate the rare side: $\sum_i p_i=20.9$ against $14$ tail resolutions. Figure~\ref{fig:empirical}(b) compares monthly tail counts with the $90\%$ interval of the independent Poisson-binomial law. No month exceeds it, the dispersion ratio across months is $0.87$, and the election month of November $2024$ ($118$ long shots) produced none. Replaying \name{} on each monthly book at its capacity boundary ($q_i=1$, $K=f(q)$) yields no insolvency and no \ac{LP} loss in $30$ months, with a median \ac{LP} profit of $1.46K$. At quarterly aggregation, one of the $11$ quarters (the third of $2026$, with $4$ tail resolutions against $1.3$ expected) exceeds its interval and yields an \ac{LP} loss with neither insolvency nor haircut, a rate consistent with a $90\%$ interval and with the \ac{LP}-loss level $\sigma_2=0.10$. Thirty months of resolutions across sports, politics, culture, and business do not contradict independence.

\section{Discussion}
\label{sec:discussion}

\noindent\textit{Dependence input.} All guarantees are conditional on the supplied joint law, and \S~\ref{sec:numerical} shows that understating dependence is the costly direction. Rare events from unrelated domains (temperature records, elections, sports upsets) have no obvious common driver (Appendix~\ref{app:selection}), and $30$ months of Polymarket resolutions do not contradict independence (\S~\ref{sec:empirical}). A conservative input costs capacity but keeps leverage: at $(n,p)=(100,0.02)$ and $\rho=0.10$, one reserve unit still backs $6.01$ units, about six times full collateral.

\noindent\textit{Scope of the venue.} The venue is deliberately narrow: it never buys claims back (Assumption~3) and may haircut settlement with probability at most $\sigma_1$ (\S~\ref{sec:basic-properties}), so it serves block-sized tail exposure while fine-priced retail trading stays on the primary venue. Sections~\ref{sec:threat} and~\ref{sec:listing} treat trust, adverse selection, manipulation, and listing.

\noindent\textit{Future work.} The first extension is the input: language models, which approach the accuracy of human forecasting crowds \cite{halawi2024approaching}, could read event descriptions to propose a book's dependence structure, alongside learned joint models \cite{rana2026parlaymarket}. The second is dynamics: \ac{LP} entry and exit with share pricing, secondary transfer of claims, and reuse of capacity freed by early resolutions (Assumption~1). The third is trader behavior, from strategic order timing and splitting to informed flow the input could learn from. The exact growth rate of $U^*$ and sharper bounds under structured dependence are open.

\section{Conclusion}
\label{sec:conclusion}
\name{} treats tail liquidity as a capital-adequacy problem: one shared reserve underwrites single-market claims, priced by \ac{CVaR} at level $\sigma_2$ and admitted by a \ac{CVaR} test at the stricter level $\sigma_1$, the frontier of its two protections. Its guarantees hold at every accepted state and arrival order, independent markets certify $\Omega(\sqrt n)$ notional per reserve unit with vanishing markup, and on Polymarket, outside the most liquid quartile, \name{} is the cheaper or only venue for block-sized long-shot orders.

\bibliographystyle{splncs04}
\bibliography{reference}
\clearpage
\appendix

\section{Omitted Proofs}
\label{app:proofs}

Table~\ref{tab:notation} summarizes the notation used in the paper and in the proofs below.

\begin{table}[!ht]
\caption{Notation. Objects marked $\ast$ are derived from $(\PX,K,\sigma_1,\sigma_2)$ in \S\S~\ref{sec:portfolio-underwriting} and~\ref{sec:capital-efficiency}.}
\label{tab:notation}
\centering
\small
\setlength{\tabcolsep}{4pt}
\begin{tabular}{ll@{\quad}ll}
\toprule
$n$ & number of markets & $X_i$, $p_i$ & outcome, YES probability\\
$\PX$ & joint law of $X$ (input) & $q$, $\mathbf{1}^\top q$ & state, total notional\\
$K$ & reserve capital & $L(q)$ & gross payout $q\cdot X$\\
$C$, $\tau$ & cost function, acceptance rule & $W(q,X)$ & terminal reserve wealth\\
$\sigma_1<\sigma_2$ & tail levels & $\CVaR_\sigma$, $\VaR_\sigma$ & upper-tail risk measures\\
$C^*$ $\ast$ & least premium & $f$ $\ast$ & tail gap\\
$\mathcal{Q}(K)$ $\ast$ & acceptance set & $U^*$, $m(\PX)$ $\ast$ & utilization, minimal tail gap\\
\bottomrule
\end{tabular}
\end{table}

\subsection{Proof of Proposition~\ref{prop:barrier}}

There is a positive-probability outcome when every sold market resolves YES. Since never-default holds, the mechanism is solvent at such an outcome. There every sold claim pays, so $q\cdot X=\mathbf{1}^\top q$ and $C(q)+K\geq\mathbf{1}^\top q$. With $C(q)\leq\sum_i\pi_i q_i$ this gives $K\geq\mathbf{1}^\top q-\sum_i\pi_i q_i=\sum_i(1-\pi_i)q_i$. \qed

\subsection{Proof of Proposition~\ref{prop:validity}}

The map $q\mapsto q\cdot X$ is linear, and \ac{CVaR} is convex as a functional of
the payout random variable, so $C^*$ is convex in $q$; also
$C^*(0)=\CVaR_{\sigma_2}^{\PX}(0)=0$, and path independence holds because
$C^*$ is a single-valued function of the state. For the price bounds, fix
$q$, $i$, $\delta>0$ and let $L=q\cdot X$. Since $X_i\in\{0,1\}$,
$L\leq L+\delta X_i\leq L+\delta$, so monotonicity and translation
invariance of \ac{CVaR} give
\[
\CVaR_{\sigma_2}^{\PX}(L)
\leq
\CVaR_{\sigma_2}^{\PX}(L+\delta X_i)
\leq
\CVaR_{\sigma_2}^{\PX}(L)+\delta ,
\]
which is $0\leq C^*(q+\delta e_i)-C^*(q)\leq\delta$. For the second claim, \eqref{P2'} holds with equality at every state because $C^*(q)=\CVaR_{\sigma_2}^{\PX}(L(q))$. If $q\in\mathcal{Q}(K)$, then $\CVaR_{\sigma_1}^{\PX}(L(q))=C^*(q)+f(q)\leq C^*(q)+K$, which is \eqref{P1'} with $C=C^*$. Both chance constraints then follow from the implication $\CVaR_\sigma^{\PX}(Y)\leq a\Rightarrow\PX[Y>a]\leq\sigma$ of \S~\ref{sec:portfolio-underwriting}. \qed

\subsection{Proof of Proposition~\ref{prop:dual}}

The premium admits the following robust-pricing dual, used in \S~\ref{sec:basic-properties} and Appendix~\ref{app:audit}.

\begin{proposition}[Dual representation]
\label{prop:dual}
For $\sigma\in(0,1)$ let $\mathcal{D}_\sigma$ be the set of probability measures $Q$ on $\{0,1\}^n$ with $Q(\omega)\leq\PX(\omega)/\sigma$ for all $\omega$. Then for every $q\in\R_{\geq0}^n$,
\[
C^*(q)=\max_{Q\in\mathcal{D}_{\sigma_2}}\E_Q[q\cdot X], \qquad f(q)=\max_{Q\in\mathcal{D}_{\sigma_1}}\E_Q[q\cdot X] -\max_{Q\in\mathcal{D}_{\sigma_2}}\E_Q[q\cdot X]\;\geq\;0 .
\]
\looseness=-1 If $C^*(q+\delta e_i)=C^*(q)$ for some $\delta>0$, every maximizer $Q$ in the $\sigma_2$-dual at $q$ satisfies $\E_Q[X_i]=0$.
\end{proposition}

The flat-increment claim, which the zero-price audit of Appendix~\ref{app:audit} uses, is proved below together with the full equivalence.

Fix $L=q\cdot X\geq0$ and $\sigma\in(0,1)$, and consider the finite linear
program
\[
\max_Q\ \sum_\omega Q(\omega)L(\omega)
\quad\text{s.t.}\quad
0\leq Q(\omega)\leq \PX(\omega)/\sigma\ \ \forall\omega,
\qquad
\sum_\omega Q(\omega)=1 .
\]
It is feasible, since $Q=\PX$ satisfies the caps for $\sigma<1$. Introduce a
free multiplier $\eta$ for the normalization and $\mu(\omega)\geq0$ for the
caps. The Lagrangian dual is
\[
\min_{\eta\in\R,\ \mu\geq0}\ \eta+\frac{1}{\sigma}\sum_\omega
\PX(\omega)\,\mu(\omega)
\quad\text{s.t.}\quad
\mu(\omega)\geq L(\omega)-\eta\ \ \forall\omega .
\]
The inner minimization sets $\mu(\omega)=(L(\omega)-\eta)_+$, so the dual
value is
$\min_{\eta}\{\eta+\frac1\sigma\E_{\PX}[(L-\eta)_+]\}=\CVaR_\sigma^{\PX}(L)$
by the variational form. Strong duality holds for feasible bounded finite
linear programs, which proves both displays; $f\geq0$ follows from
$\mathcal{D}_{\sigma_2}\subseteq\mathcal{D}_{\sigma_1}$.

For the flat-increment case: ($\Leftarrow$) if $Q_0$ is optimal
at $q+\delta e_i$ with $\E_{Q_0}[X_i]=0$, then
\[
C^*(q+\delta e_i)
=\E_{Q_0}[(q+\delta e_i)\cdot X]
=\E_{Q_0}[q\cdot X]
\leq C^*(q)\leq C^*(q+\delta e_i),
\]
the last step by monotonicity, so equality holds and $Q_0$ is
optimal at $q$. ($\Rightarrow$) Let $Q_0$ be any maximizer at $q$,
which exists since the program is a finite LP. Then
\[
C^*(q+\delta e_i)\;\geq\;\E_{Q_0}[(q+\delta e_i)\cdot X]
\;=\;C^*(q)+\delta\,\E_{Q_0}[X_i].
\]
Flatness forces $\E_{Q_0}[X_i]\leq0$, hence $=0$, and then $Q_0$ attains
$C^*(q+\delta e_i)$, so it is optimal at both states. We note that an
arbitrary maximizer at $q+\delta e_i$ need not have zero $X_i$-mass: with
$n=2$, $\PX(1,0)=\PX(0,1)=1/2$, $\sigma=1/2$, $q=e_1$, $\delta=1$, and $i=2$, the
increment is flat while the point mass on $(0,1)$ is optimal at $(1,1)$
and assigns $X_2$ mass one. \qed

\subsection{Proof of Theorem~\ref{thm:ratio-form}}

\ac{CVaR} is positively homogeneous, so $f(\lambda q)=\lambda f(q)$ for every
$\lambda\geq0$; hence $\mathcal{Q}(K)=K\,\mathcal{Q}(1)$ and the utilization
in \eqref{eq:capital-utilization} does not depend on $K$. For nonzero
$q\geq0$, write $q=(\mathbf{1}^\top q)u$ with
$u=q/(\mathbf{1}^\top q)\in\Delta_n$; homogeneity gives
$f(q)=(\mathbf{1}^\top q)f(u)$, so the constraint $f(q)\leq K$ reads
$\mathbf{1}^\top q\leq K/f(u)$ whenever $f(u)>0$. Maximizing over directions,
$U^*(K)=\sup_{u\in\Delta_n}1/f(u)=1/\min_{u\in\Delta_n}f(u)$. Continuity of
\ac{CVaR} in $u$ and compactness of $\Delta_n$ imply the minimum is attained; if
it is zero, scaling a zero-gap minimizer keeps the state accepted at all
scales, so utilization is unbounded. \qed

\subsection{Proof of Corollary~\ref{cor:capacity}}

Call a trade sequence \emph{admissible} if every prefix state is accepted.

\emph{Increment lemma.} For every $q\in\R^n_{\geq0}$, $i\in[n]$,
$\delta>0$: $f(q+\delta e_i)\leq f(q)+\delta$. Indeed, with $L=q\cdot X$ we
have $L\leq L+\delta X_i\leq L+\delta$ pointwise, so monotonicity and
translation invariance give
$\CVaR_{\sigma_1}(L+\delta X_i)\leq\CVaR_{\sigma_1}(L)+\delta$, while
monotonicity gives
$\CVaR_{\sigma_2}(L+\delta X_i)\geq\CVaR_{\sigma_2}(L)$; subtract. (Note
$f$ itself need not be monotone; only this upper bound is used.)

\emph{Upper bound.} Every prefix of an admissible sequence lies in
$\mathcal{Q}(K)$; for $q\neq0$, writing $q=(\mathbf{1}^\top q)u$ with
$u\in\Delta_n$, homogeneity gives
$f(q)=(\mathbf{1}^\top q)f(u)\geq(\mathbf{1}^\top q)\,m$, hence
$\mathbf{1}^\top q\leq K/m=U^*K$.

\emph{Approachability.} Let $u^*$ attain $m$
(Theorem~\ref{thm:ratio-form}) and set
$q^\varepsilon=(1-\varepsilon)(K/m)u^*$, so that
$f(q^\varepsilon)=(1-\varepsilon)K$ and
$\mathbf{1}^\top q^\varepsilon=(1-\varepsilon)U^*K$. Choose an integer
$R\geq\mathbf{1}^\top q^\varepsilon/(\varepsilon K)$ and buy, in $R$
rounds, $q^\varepsilon_i/R$ units of each market $i$ with
$q^\varepsilon_i>0$, in any order within rounds. A generic prefix state is
$q_{\mathrm{pre}}=\frac{r}{R}q^\varepsilon
+\sum_{i\in A}\frac{q^\varepsilon_i}{R}e_i$ for some $r<R$ and
$A\subseteq[n]$. Applying the increment lemma once per element of $A$,
homogeneity, and $\sum_{i\in A}q^\varepsilon_i\leq\mathbf{1}^\top q^\varepsilon$,
\[
f(q_{\mathrm{pre}})
\;\leq\;\frac{r}{R}\,(1-\varepsilon)K+\frac{\mathbf{1}^\top
q^\varepsilon}{R}
\;\leq\;(1-\varepsilon)K+\varepsilon K\;=\;K,
\]
so every prefix is accepted, and each trade has size
$q^\varepsilon_i/R\leq\varepsilon K$. With bundle trades, the single
proposed state $q^*=(K/m)u^*$ satisfies $f(q^*)=K$ and is accepted,
attaining $U^*K$ exactly. \qed

\subsection{Proof of Corollary~\ref{cor:iid-symmetric} (asymptotics)}

Normalize $Z_n:=(S_n-np)/\sqrt{np(1-p)}$. The central limit theorem gives
$Z_n\Rightarrow Z\sim\mathcal{N}(0,1)$, and the normalized binomial
sequence is uniformly integrable since $\sup_n\E[Z_n^2]=1$. At fixed tail levels, \ac{CVaR} is continuous
under this convergence, so
$\CVaR_\sigma(Z_n)\to\CVaR_\sigma(Z)=\varphi(z_\sigma)/\sigma$, using the
Gaussian formula $\CVaR_\sigma(\mu+sZ)=\mu+s\,\varphi(z_\sigma)/\sigma$.
By translation invariance and positive homogeneity,
\[
\CVaR_\sigma(S_n)
=np+\sqrt{np(1-p)}\,
\Bigl[\frac{\varphi(z_\sigma)}{\sigma}+o(1)\Bigr].
\]
Subtracting the two tail levels cancels the mean term,
\[
\CVaR_{\sigma_1}(S_n)-\CVaR_{\sigma_2}(S_n)
=\sqrt{np(1-p)}\,[c(\sigma_1,\sigma_2)+o(1)],
\]
and substituting into \eqref{eq:symmetric-utilization} proves
\eqref{eq:symmetric-sqrt-n}; since upper-tail Gaussian CVaR is strictly
decreasing in the level, $c(\sigma_1,\sigma_2)>0$. \qed

\subsection{The CVaR Perturbation Lemma}

\begin{lemma}[CVaR perturbation under total variation]
\label{lem:cvar-tv}
Let $L\in[0,M]$ under two probability measures $\mathbb{P}$ and $\mathbb{Q}$.
For any $\sigma\in(0,1)$,
\[
\left|
\CVaR_\sigma^{\mathbb{P}}(L)
-
\CVaR_\sigma^{\mathbb{Q}}(L)
\right|
\leq
\frac{M}{\sigma}\TV(\mathbb{P},\mathbb{Q}).
\]
\end{lemma}

Using the variational representation, choose an optimizer
$\eta_{\mathbb{Q}}$ for the $\mathbb{Q}$ objective; since $L\in[0,M]$, it
can be taken in $[0,M]$. Evaluating the $\mathbb{P}$ objective at
$\eta_{\mathbb{Q}}$ and subtracting the $\mathbb{Q}$ optimum,
\[
\CVaR_\sigma^{\mathbb{P}}(L)-\CVaR_\sigma^{\mathbb{Q}}(L)
\;\leq\;
\frac{1}{\sigma}
\Bigl(\E_{\mathbb{P}}[(L-\eta_{\mathbb{Q}})_+]
-\E_{\mathbb{Q}}[(L-\eta_{\mathbb{Q}})_+]\Bigr).
\]
The function $(L-\eta_{\mathbb{Q}})_+$ is bounded with range contained in
an interval of length at most $M$, and for any such function $g$, centering
at the midpoint of its range and using
$\TV(\mathbb{P},\mathbb{Q})=\tfrac12\lVert\mathbb{P}-\mathbb{Q}\rVert_1$
gives the oscillation bound
$|\E_{\mathbb{P}}[g]-\E_{\mathbb{Q}}[g]|\leq M\,\TV(\mathbb{P},\mathbb{Q})$.
Hence the difference is at most $M\,\TV(\mathbb{P},\mathbb{Q})/\sigma$;
swapping $\mathbb{P}$ and $\mathbb{Q}$ gives the absolute-value bound. \qed

\subsection{Proof of Theorem~\ref{thm:tv-robustness}}

For every $u\in\Delta_n$ the loss $u\cdot X$ lies in $[0,1]$, so applying
Lemma~\ref{lem:cvar-tv} at the two tail levels gives
$|f^{\PX}(u)-f^{\PX^0}(u)|\leq\epsilon(1/\sigma_1+1/\sigma_2)=\alpha$
uniformly in $u$. Taking minima over the simplex preserves a uniform error
bound, so $|m(\PX)-m_0|\leq\alpha$. If $\alpha<m_0$, then $m(\PX)>0$ and
Theorem~\ref{thm:ratio-form} gives $U^*(\mathbb{P})=1/m(\mathbb{P})$ for
both distributions; the displayed bounds follow by inverting
$m_0-\alpha\leq m(\PX)\leq m_0+\alpha$. The final relative-error inequality
uses
$|U^*(\PX)/U^*(\PX^0)-1|=|m(\PX)-m_0|/m(\PX)\leq\alpha/(m_0-\alpha)
\leq2\alpha/m_0$ for $\alpha\leq m_0/2$, and $\sigma_1<\sigma_2$. \qed

\subsection{Proof of Corollary~\ref{cor:default-robust}}

Proposition~\ref{prop:validity} gives $\PX[A]\leq\sigma_1$ for $A=\{L(q)>C^*(q)+K\}$ at every accepted state, and $\mathbb{Q}[A]\leq\PX[A]+\TV(\PX,\mathbb{Q})$ by the definition of total variation. The second claim repeats the argument with \eqref{P2}. \qed

\subsection{Proof of Proposition~\ref{prop:block-cutoff}}

All-primary execution costs $P_i(\delta)$, while a split of size $x$ costs
$\mathcal{E}_i(q,\delta,x)=C^*(q+xe_i)-C^*(q)+P_i(\delta-x)$; weak improvement
is $\mathcal{E}_i(q,\delta,x)\leq P_i(\delta)$, which rearranges to
\eqref{eq:block-routing} after dividing by $x>0$, with saving
$P_i(\delta)-\mathcal{E}_i(q,\delta,x)$ under strict inequality. The premium
form subtracts $p_i$ from \eqref{eq:block-routing}. If
$P_i(s)=(p_i+\kappa_i)s$, then
$\bar\pi^P_i(\delta,x)=p_i+\kappa_i$, so the routing condition becomes
$\kappa_i\geq\pi_{\mathrm{PU},i}(q,x)-p_i$; under $\kappa_i=rT_i$ with $r>0$ this is
$T_i\geq(\pi_{\mathrm{PU},i}(q,x)-p_i)_+/r$, the positive part covering the
case $\pi_{\mathrm{PU},i}(q,x)\leq p_i$ in which every $T_i\geq0$ qualifies.
At the empty book, positive homogeneity gives
$C^*(xe_i)=x\,\CVaR_{\sigma_2}(X_i)$, and the variational form evaluates the
Bernoulli \ac{CVaR}: for $0<p_i\leq\sigma_2$ the objective
$\eta+\frac{1}{\sigma_2}\E[(X_i-\eta)_+]$ has derivative
$1-p_i/\sigma_2\geq0$ on $[0,1]$, so the minimum is at $\eta=0$ with value
$p_i/\sigma_2$. Monotonicity in $x$ is the monotonicity of difference
quotients of the convex map $t\mapsto C^*(q+te_i)$; monotonicity in the
base point is the convexity inequality
$C^*(q+te_i+xe_i)-C^*(q+te_i)\geq C^*(q+xe_i)-C^*(q)$ for
$t>0$. \qed

\subsection{Proof of Proposition~\ref{prop:lp-apr}}

Let $L=q\cdot X$. Since $C^*(q)=\CVaR_{\sigma_2}(L)$,
\[
\E[\Pi_L(q,X)]
=C^*(q)-\E[L]
=\CVaR_{\sigma_2}(L)-\E[L].
\]
The last quantity is nonnegative because
$\PX\in\mathcal{D}_{\sigma_2}$ (Proposition~\ref{prop:dual}). If $L$ is
$\PX$-a.s.\ constant, \ac{CVaR} equals the mean by translation invariance. If not,
pick $t$ with $\alpha:=\PX[L>t]\in(0,1)$, set
$c=\min(1/\sigma_2,\,1/\alpha)>1$ and
$d=(1-c\alpha)/(1-\alpha)\in[0,1)$, and define $Q=c\,\PX$ on $\{L>t\}$ and
$Q=d\,\PX$ on $\{L\leq t\}$. Then $Q$ is a probability measure in
$\mathcal{D}_{\sigma_2}$, and using $(d-1)(1-\alpha)=-(c-1)\alpha$,
\[
\E_Q[L]-\E[L]
=(c-1)\,\alpha\,\bigl(\E[L\mid L>t]-\E[L\mid L\leq t]\bigr)\;>\;0,
\]
since $\E[L\mid L>t]>t\geq\E[L\mid L\leq t]$, $c>1$, and $\alpha>0$. Hence
$\CVaR_{\sigma_2}(L)\geq\E_Q[L]>\E[L]$. Under full support and $q\neq0$,
$L$ takes the value $0$ (all-NO outcome) and a positive value (any outcome
with $X_j=1$ for some $j\in\operatorname{supp}(q)$), each with positive
probability, so $L$ is non-degenerate.

Now take $q_{\phi,u}=\phi K u/f(u)$. Positive homogeneity gives
\[
C^*(q_{\phi,u})-\E[q_{\phi,u}\cdot X]
=
\phi\,\frac{K}{f(u)}
\left(\CVaR_{\sigma_2}(u\cdot X)-\E[u\cdot X]\right),
\]
which is the expected-profit formula after substituting $a(u)$. \qed

\paragraph{Conditional APR.}
For an exogenous holding horizon $H>0$ years, define the realized simple
\ac{APR} as $\mathrm{APR}_L(q,X;H):=\Pi_L(q,X)/(KH)$. Dividing the ray formula
by $KH$ gives the conditional expected \ac{APR}
\[
\E_{\PX}[\mathrm{APR}_L(q_{\phi,u},X;H)]
=\frac{\phi}{H}\,\frac{a(u)}{f(u)} .
\]
In the i.i.d.\ benchmark with $u^{\mathrm{sym}}=\mathbf{1}/n$ and
$S_n=\sum_iX_i$, $u^{\mathrm{sym}}\cdot X=S_n/n$, and positive homogeneity
cancels the common factor $1/n$, yielding
\[
\E_{\PX}[\mathrm{APR}_L(q_{\phi,u^{\mathrm{sym}}},X;H)]
=\frac{\phi}{H}\,
\frac{\CVaR_{\sigma_2}^{\PX}(S_n)-np}
{\CVaR_{\sigma_1}^{\PX}(S_n)-\CVaR_{\sigma_2}^{\PX}(S_n)} .
\]

\section{Axiom Audit: Details}
\label{app:audit}

This appendix details the axiom audit of \S~\ref{sec:basic-properties}. It maps \name{} onto the axiom systems of \cite{frongillo2024axiomatic,angeris2024geometry}, shows that $C^*$ restricts a classical coherent-risk cost function, and bounds the zero-price edge.

\paragraph{Axiom mapping.}
\name{} keeps path independence (Axioms~2 and~5 of \cite{frongillo2024axiomatic}), no-arbitrage (Axiom~1, NoDominatedTrades, via the price bounds of Proposition~\ref{prop:validity}), and demand responsiveness (Axiom~4, by convexity of $C^*$), with inventory dependence across portfolio directions through $f$. It relinquishes always-quoting (Axiom~3, Liquidation, of \cite{frongillo2024axiomatic}, and upward closedness of the reachable set in \cite{angeris2024geometry}), since Theorem~\ref{thm:frontier} makes refusal beyond $\mathcal{Q}(K)$ the price of least-premium quoting. It replaces the never-debt condition $S\subseteq\R^n_+$ of \cite{angeris2024geometry} by \ac{CVaR} solvency with a proportional haircut, the source of leverage (Proposition~\ref{prop:barrier} versus Corollary~\ref{cor:iid-symmetric}), and it does not assume their convex-reachable-set condition: $\mathcal{Q}(K)$ is closed and star-shaped but need not be convex, and no result uses convexity. The induced scoring rule is proper but not strictly proper. A convex cost function scores reports through its convex conjugate \cite{abernethy2013efficient}, and the conjugate of a positively homogeneous cost function is an indicator, here of the moment set $\mathcal{M}_{\sigma_2}:=\{\E_Q[X]:Q\in\mathcal{D}_{\sigma_2}\}\subseteq[0,1]^n$ (Proposition~\ref{prop:dual}, via the complete-space lift below), so all reports in $\mathcal{M}_{\sigma_2}$ receive identical scores. \name{} is therefore a tail-liquidity backstop rather than a price-discovery venue, while the execution incentives of price-taking traders are unaffected (\S~\ref{sec:application}).

\paragraph{$C^*$ restricts a classical cost function.}
Lift the $n$ non-exclusive markets to the complete outcome space
$\{0,1\}^n$ with its $2^n$ Arrow--Debreu securities, and let
$\widetilde{C}(z)=\CVaR_{\sigma_2}^{\PX}(\text{payout of }z)$ be the cost of
an inventory $z$ over those securities. Then $\widetilde{C}$ is convex,
nondecreasing, positively homogeneous, and translation-invariant
($\widetilde{C}(z+c\mathbf{1})=\widetilde{C}(z)+c$, since the complete-space
grand bundle pays $1$ in every state), so it is a classical coherent-risk
market-making cost function \cite{artzner1999coherent,othman2011liquidity},
meeting the cost-function requirements of \cite{frongillo2024axiomatic} except that it is only weakly increasing, and the path-independence, information-incorporation, and no-arbitrage conditions of \cite{abernethy2013efficient} but not their differentiability or expressiveness. The flat directions this permits are exactly the zero-price edge treated below. \name{} is the
restriction of $\widetilde{C}$ to the cone of single-market YES inventories
($q\geq0$), together with the acceptance cap. The novelty therefore lies in (i) the two-level split that funds
the $\sigma_2$ tail by premiums and only the $\sigma_1$--$\sigma_2$ gap by
capital, (ii) the resulting acceptance rule, and (iii) the
capital-utilization theory those induce. Translation invariance holds on the
complete space but not on the single-market cone: the $n$ YES securities are
non-exclusive, so their grand bundle does not pay a constant and cannot serve
as cash.

\paragraph{Flat increments are bounded.}
By Proposition~\ref{prop:dual}, a flat increment $C^*(q+\delta e_i)=C^*(q)$
forces every $\sigma_2$-dual maximizer at the pre-trade state $q$ to satisfy
$\E_Q[X_i]=0$. The proof in Appendix~\ref{app:proofs} gives the full equivalence. No free claims exist at the origin, since
$C^*(\delta e_i)=\delta\min(1,p_i/\sigma_2)>0$ for $p_i>0$. Moreover, if
$C^*(q+\delta e_i)=C^*(q)$ with $0<p_i\leq\sigma_2$, then monotonicity gives
$\delta\leq\sigma_2\,C^*(q)/p_i$, so the flat range is finite. A zero price
does not exempt a trade from the acceptance check, and \eqref{P1'}, \eqref{P2'} hold at
any accepted post-trade state. An implementation may overlay a quote floor as a fee, but no result requires it.

\paragraph{Not a special case of liquidity-based \acp{AMM}.}
The liquidity-based \ac{AMM} class of \cite{amini2025decentralized} requires a strictly increasing utility, which $-\CVaR_{\sigma_2}$ violates because it ignores wealth gains outside the $\sigma_2$ tail. That class is also liquidity-bounded-loss, which is the never-default condition under which Proposition~\ref{prop:barrier} caps capacity at $O(K)$ whenever unit prices stay below some $\bar\pi<1$. This second departure is what buys leverage: \name{} reaches $\Omega(\sqrt n)\,K$ in the i.i.d.\ benchmark (Corollary~\ref{cor:iid-symmetric}) precisely by settling the deep tail with a haircut instead of reserving for it.

\section{Market Selection for the Underwriting Book}
\label{app:selection}

Which markets to list is an operator decision outside the mechanism, but the
mechanism scores any candidate book exactly: certified capacity is
$U^*(K)\,K=K/m(\PX)$ (Theorem~\ref{thm:ratio-form}), computable from the
scalar law of $u\cdot X$. This appendix collects what the paper's results
imply for that choice.

\begin{lemma}[Listing is quote-neutral and capacity-monotone]
\label{lem:listing}
Extend the book by an $(n{+}1)$st market, supplying a joint law on
$\{0,1\}^{n+1}$ whose marginal on the first $n$ coordinates is $\PX$, and
write $C^{*\prime}$, $f'$, $\mathcal{Q}'(K)$, $U^{*\prime}(K)$ for the
extended mechanism. Then $C^{*\prime}(q,0)=C^*(q)$ and $f'(q,0)=f(q)$ for
every old state $q$, so quotes and acceptance decisions along trade paths in the old markets are unchanged. Moreover, $\mathcal{Q}(K)\times\{0\}\subseteq \mathcal{Q}'(K)$ and $U^{*\prime}(K)\geq U^*(K)$ for every $K>0$.
\end{lemma}

\begin{proof}
$(q,0)\cdot X'=q\cdot(X_1,\dots,X_n)$ has the same law under the extended
joint law as $q\cdot X$ under $\PX$ by marginal consistency, and $C^*$ and
$f$ depend on the state only through the law of this scalar loss. The
capacity claim takes the supremum in \eqref{eq:capital-utilization} over a
superset.
\end{proof}

Listing is thus free at the certificate level: an unused listing changes no quote and no acceptance decision, and never lowers capacity. The cost sits on the input side: the operator must supply a joint law over a larger book (\S~\ref{sec:threat}). Neutrality holds at zero new exposure, and once trades arrive on the new market, portfolio pricing moves old quotes,
which is the design.

Capacity comes from a thin joint upper tail rather than from marginal rarity alone. In the dependence experiment of \S~\ref{sec:numerics}
the marginals are held at $p=0.02$ while pairwise correlation rises: the
certificate falls from $55.34$ under independence to $6.01$ at $\rho=0.10$
and $1.67$ at $\rho=0.50$, against the comonotone endpoint $1.25$. Rarity is unchanged, so the collapse is due entirely to dependence. The worked examples of
Appendix~\ref{app:numerics} make the same point: $100$ markets
forming two independent comonotone blocks certify $U=8.00$, exactly the value
of two independent markets, against $39.26$ for $100$ i.i.d.\ markets
(Table~\ref{tab:app-examples}, $p=0.05$). Diversification counts independent clusters rather than raw markets. Near-independence is provably
benign: a book within \ac{TV} distance $\epsilon$ of the independent product
keeps $m$ within $\epsilon(1/\sigma_1+1/\sigma_2)$
(Theorem~\ref{thm:tv-robustness}).

Two practical rules follow. First, screen candidates by their effect on the
joint tail rather than by rarity: in the families computed above, additions
strongly dependent on the current book add little certified capacity and
weakly dependent ones add much more, and any specific comparison is
computable exactly, since a candidate book is scored by $K/m$. Second,
selection need not be solved in advance: the acceptance rule prices listings
online, and Lemma~\ref{lem:listing} makes trial listings safe.

\section{Operational Note: Cold-Start Pricing and the Diversified Seed}
\label{app:bootstrap}

This appendix is operational and changes nothing in the mechanism: the initial
state is $q=0$ in the main text, and $C^*,\tau^*$ are untouched. It addresses
the price an early trader sees on a thin book and
quantifies a standard remedy in exact Binomial arithmetic.

\paragraph{The cold start.}
On an empty book the marginal quote for the first $\delta$ units of YES on
market $i$ is $C^*(\delta e_i)/\delta=\min(1,p_i/\sigma_2)$, because the loss
$\delta X_i$ is a single Bernoulli variable whose $\sigma_2$-\ac{CVaR} is
$\delta\,p_i/\sigma_2$ when $p_i\le\sigma_2$. For a long shot this exceeds fair value: at $p_i=0.02$, $\sigma_2=0.10$ the unit quote is $0.20$, ten
times $p_i$. The premium reflects the undiversified tail risk of an isolated claim, so single claims are better routed to the primary venue.

\paragraph{The diversified seed.}
The marginal quote on market $i$ at state $q$ is the tail-tilted probability $\E_{Q^*}[X_i]$ for a maximizer $Q^*$ of the $\sigma_2$-dual at $q$ (Proposition~\ref{prop:dual}), the largest such value when $Q^*$ is not unique. On a
diversified book this tilt is small and prices fall toward fair value. An
operator can therefore initialize the book at a symmetric seed
$q_0=(s,\dots,s)$ funded from $K$. Two properties hold. First, by positive homogeneity the seeded marginal price is \emph{scale-invariant} in the
seed size $s$, so an arbitrarily small seed unlocks the diversified price while $s$ sets only the \emph{depth} over which it is offered. The seed plays the role
of the liquidity parameter $b$ in \ac{LMSR} \cite{hanson2007lmsr}, a bounded
liquidity subsidy that buys depth. Second, in exact Binomial computation, convergence to fair value is governed by $np$, the expected number of tail events in the book, rather than by $n$ alone.

\paragraph{Numbers.}
For $K=\$1\mathrm{M}$, $p=0.02$, $(\sigma_1,\sigma_2)=(0.01,0.10)$, the
empty-book single-claim quote is $0.20$ ($10\times$ fair) for every $n$. The
seeded marginal price falls with $n$:
\begin{center}
\begin{tabular}{lcccccc}
\toprule
$n$ & $10$ & $30$ & $100$ & $300$ & $1000$ & $3000$\\
\midrule
seeded marginal & $0.200$ & $0.089$ & $0.064$ & $0.050$ & $0.031$ & $0.027$\\
$\times$ fair & $10.0$ & $4.4$ & $3.2$ & $2.5$ & $1.5$ & $1.3$\\
\bottomrule
\end{tabular}
\end{center}
For small $np$ the $\sigma_2$-tail is still a single-market event, so at $n=10$ ($np=0.2$) the marginal stays at $p/\sigma_2$. At $n=100$ the seeded marginal ($3.2\times$ fair) exceeds the average boundary price of Table~\ref{tab:num-capacity-price} ($2.36\times$ fair) because the Binomial \ac{CVaR} has kinks, and a single-market increment pays the one-sided derivative, which at a kink exceeds the average price along the symmetric ray. The seed's cost scales with provided depth, while the
per-depth cost falls with the size of the diversified pool.

\paragraph{Scope.}
Seeding is an \ac{LP}-side initialization lever outside the core mechanism. Its exact cost depends on whether it is run as a pure liquidity subsidy or a self-financed
position. A richer alternative, hedging the reserve against the primary order
book, is cross-venue and left to future work.

\section{Supplementary Experiments}
\label{app:numerics}

This appendix gives the data rules of the empirical experiments and the remaining numerical experiments. All capacity and acceptance computations in this paper use exact Binomial evaluation or exact finite-law bucket convolution, and sample-average approximation appears only as a comparison baseline below.

\paragraph{Polymarket data.} Both empirical experiments of \S~\ref{sec:numerics} use only public market data. \emph{Order books.} On 2026-09-24 04:04 UTC we listed all active markets (six listing orders, union of $11{,}038$ markets), kept binary Yes/No markets whose long-shot side (the side priced below one half, Assumption~3) is priced in $[0.005,0.05]$ with at least $14$ days to resolution, and fetched the YES book of each. The CLOB book already mirrors complementary NO orders at $1-p$, so it carries the executable depth of both sides: a YES long shot is bought on the YES asks, and a NO long shot on the YES bids at $1-p$. The NO book is used only as a consistency check. Depth and execution prices are computed by walking the long-shot side's ask ladder. The \name{} book takes $p_i$ equal to the mid, the independent product as $\PX$, one market per event (the most liquid), $K=\$100\mathrm{K}$, and the symmetric state at $\phi=0.5$, and a single-market tranche is priced exactly by convolving the other markets' Bernoulli laws. Liquidity tiers use the platform's reported resting liquidity. \emph{Repeated snapshots.} We repeated the full snapshot nine more times on 2026-09-24, ten snapshots in all, and every number in \S~\ref{sec:empirical} is from the first. Across the ten, the number of valid long-shot books ranges from $620$ to $673$, the most liquid tier fills 10K shares at $1.11$ to $1.12$ times the mid, the median cost of 10K shares outside it stays above $4.5$ times the mid, and no market outside it fills 100K shares below twice the mid. The one-per-event book has $293$ to $311$ markets with $\sum_i p_i$ between $7.9$ and $8.3$, and its symmetric-ray premium stays between $1.63$ and $1.66$ times the mean mid. Outside the most liquid quartile, \name{} is cheaper or the only venue in $64\%$ to $76\%$ of markets at 10K shares, $88\%$ to $95\%$ at 50K, and $98\%$ to $100\%$ at 100K, and the basket saves $8\%$ to $20\%$ at 10K shares and $20\%$ to $29\%$ at 50K, while the primary book cannot fill a 250K-share basket in any snapshot.

\begin{figure}[t]
\centering
\includegraphics[width=.7\textwidth]{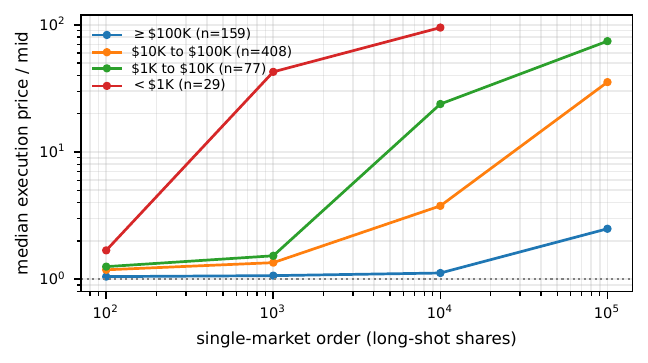}
\caption{Median execution price relative to the mid for a single-market long-shot order on $673$ live long-shot books, by resting liquidity (snapshot 2026-09-24).}
\label{fig:app-depth}
\end{figure} \emph{Resolved markets.} We enumerated events resolved between 2024-01-01 and 2026-08-31 with at least \$10K of volume, kept binary markets with a definite resolution, at least \$10K of volume, and a lifetime of at least $45$ days (at most eight per event, earliest listed), fetched daily YES prices from the platform's public price history, and set $p_i$ to the mean price over the seven days ending $30$ days before the market's end date, requiring at least three observations in that window and fifteen before it. Following Assumption~3, the long-shot side of a market is whichever of YES and NO trades below one half, and long shots are markets whose long-shot side is priced in $[0.005,0.05]$. One market per event is kept (the earliest listed, which is exogenous to the outcome, whereas selecting by lifetime volume favors eventual winners), and months with fewer than five markets are dropped, leaving $907$ markets. Category labels come from the platform's event tags.

\paragraph{Dependence misspecification is asymmetric.}
Table~\ref{tab:app-misspec} certifies the book on the symmetric ray at $(n,p)=(100,0.02)$ under a supplied beta-binomial correlation $\rho_s$, places it at the capacity boundary, and reports the exact insolvency probability under a true correlation $\rho$. Every entry with $\rho\leq\rho_s$ stays below $\sigma_1=0.01$, while understated dependence can exceed it.

\begin{table}[t]
\centering
\caption{Insolvency probability $\Pr[W<0]$ under the true correlation $\rho$ for a book certified under the supplied correlation $\rho_s$, with $(n,p)=(100,0.02)$ and $(\sigma_1,\sigma_2)=(0.01,0.10)$. Bold entries exceed $\sigma_1$.}
\label{tab:app-misspec}
\small
\setlength{\tabcolsep}{4pt}
\begin{tabular}{ccccccccc}
\toprule
 & & \multicolumn{7}{c}{true $\rho$}\\
\cmidrule(lr){3-9}
$\rho_s$ & $U_{\mathrm{sym}}$ & $0$ & $0.005$ & $0.01$ & $0.02$ & $0.05$ & $0.1$ & $0.3$\\
\midrule
$0$ & $55.34$ & $0.0041$ & $\mathbf{0.0184}$ & $\mathbf{0.0338}$ & $\mathbf{0.0579}$ & $\mathbf{0.0901}$ & $\mathbf{0.0988}$ & $\mathbf{0.0752}$\\
$0.005$ & $36.16$ & $0.0002$ & $0.0033$ & $0.0098$ & $\mathbf{0.0250}$ & $\mathbf{0.0559}$ & $\mathbf{0.0726}$ & $\mathbf{0.0646}$\\
$0.01$ & $27.00$ & $0.0000$ & $0.0005$ & $0.0027$ & $\mathbf{0.0106}$ & $\mathbf{0.0351}$ & $\mathbf{0.0542}$ & $\mathbf{0.0564}$\\
$0.02$ & $18.60$ & $0.0000$ & $0.0001$ & $0.0007$ & $0.0045$ & $\mathbf{0.0221}$ & $\mathbf{0.0410}$ & $\mathbf{0.0498}$\\
$0.03$ & $14.43$ & $0.0000$ & $0.0000$ & $0.0001$ & $0.0012$ & $\mathbf{0.0111}$ & $\mathbf{0.0272}$ & $\mathbf{0.0419}$\\
$0.05$ & $10.11$ & $0.0000$ & $0.0000$ & $0.0000$ & $0.0001$ & $0.0035$ & $\mathbf{0.0139}$ & $\mathbf{0.0320}$\\
$0.1$ & $6.01$ & $0.0000$ & $0.0000$ & $0.0000$ & $0.0000$ & $0.0003$ & $0.0036$ & $\mathbf{0.0193}$\\
\bottomrule
\end{tabular}
\end{table}

\paragraph{Default probabilities at the boundary.} Table~\ref{tab:app-default} reports, for every cell of Table~\ref{tab:num-capacity-price}, the exact supplied-law probabilities of insolvency and \ac{LP} loss at the symmetric boundary state $f(q)=K$, the expected shortfall, and the mean haircut conditional on insolvency. Every cell satisfies $\Pr[W<0]\leq\sigma_1$ and $\Pr[W<K]\leq\sigma_2$, as the \ac{CVaR} constraints imply. On Binomial tails, both probabilities sit inside their levels.

\begin{table}[t]
\centering
\caption{Supplied-law default probabilities at the boundary, $(\sigma_1,\sigma_2)=(0.01,0.10)$.}
\label{tab:app-default}
\small
\begin{tabular}{cccccccc}
\toprule
$p$ & $n$ & $U_{\mathrm{sym}}$ & avg.\ price & $\Pr[W<0]$ & $\Pr[W<K]$ & $\E[(L-C^*-K)_+]/K$ & $\E[h\mid W<0]$\\
\midrule
$0.02$ & $25$ & $22.26$ & $0.0814$ & $0.0014$ & $0.0132$ & $0.0012$ & $0.776$\\
$0.02$ & $100$ & $55.34$ & $0.0472$ & $0.0041$ & $0.0508$ & $0.0017$ & $0.900$\\
$0.02$ & $400$ & $129.61$ & $0.0331$ & $0.0034$ & $0.0327$ & $0.0015$ & $0.926$\\
$0.05$ & $25$ & $17.48$ & $0.1371$ & $0.0072$ & $0.0341$ & $0.0017$ & $0.941$\\
$0.05$ & $100$ & $39.26$ & $0.0909$ & $0.0043$ & $0.0282$ & $0.0014$ & $0.935$\\
$0.05$ & $400$ & $88.74$ & $0.0699$ & $0.0038$ & $0.0480$ & $0.0014$ & $0.953$\\
$0.10$ & $25$ & $13.95$ & $0.2175$ & $0.0023$ & $0.0334$ & $0.0013$ & $0.880$\\
$0.10$ & $100$ & $30.53$ & $0.1551$ & $0.0046$ & $0.0399$ & $0.0013$ & $0.955$\\
$0.10$ & $400$ & $66.79$ & $0.1270$ & $0.0043$ & $0.0436$ & $0.0013$ & $0.970$\\
\bottomrule
\end{tabular}
\end{table}

\paragraph{Asymmetric directions at small $n$.}
For $n\leq12$ we enumerate the joint law on $\{0,1\}^n$, check every uniform
$k$-subset direction, and run multistart SLSQP over the simplex. The symmetric
direction is beaten in 13 of 22 tested $(n,p)$ cases. Table~\ref{tab:app-nonopt}
shows representative rows. These cases are usually staircase effects: a
smaller subset can sit at a lower binomial tail gap than the full pool. Two
cases also exhibit non-uniform interior minimizers. Thus $U_{\mathrm{sym}}$ is an achievable certificate, and exact $U^*$ requires solving the simplex problem.

\begin{table}[t]
\centering
\caption{Representative exact small-$n$ minimization results,
$(\sigma_1,\sigma_2)=(0.01,0.10)$. ``Gain'' is
$U_{\mathrm{best}}/U_{\mathrm{sym}}-1$.}
\label{tab:app-nonopt}
\begin{tabular}{ccccc}
\toprule
$n$ & $p$ & $U_{\mathrm{sym}}$ & best direction & gain\\
\midrule
$3$  & $0.05$ & $4.52$  & $2$-subset & $77.0\%$\\
$4$  & $0.05$ & $4.43$  & $2$-subset & $80.7\%$\\
$8$  & $0.02$ & $8.56$  & $5$-subset & $49.1\%$\\
$9$  & $0.05$ & $8.17$  & interior, one weight $2\times$ & $11.7\%$\\
$10$ & $0.05$ & $8.91$  & interior, one weight $2\times$ & $6.9\%$\\
$12$ & $0.02$ & $13.13$ & symmetric & $0\%$\\
\bottomrule
\end{tabular}
\end{table}

\paragraph{Heterogeneous markets.}
The main i.i.d.\ benchmark is not needed for the oracle. In
Table~\ref{tab:app-heterogeneous}, $100$ independent markets are split into
five equal buckets with $p_i\in\{0.005,0.01,0.02,0.05,0.10\}$. For each
direction, the scalar loss law $u\cdot X$ is computed by exact convolution of
the five independent Binomial bucket counts. Equal exposure certifies
$44.14K$ notional, and inverse standard-deviation weighting
($u_i\propto1/\sqrt{p_i(1-p_i)}$) raises this to $57.04K$ by
allocating more mass to rarer buckets. The same \ac{CVaR} machinery thus handles heterogeneous marginals, and capacity depends on the exposure choice.

\begin{table}[t]
\centering
\caption{Exact bucket-convolution capacities for heterogeneous independent
markets.}
\label{tab:app-heterogeneous}
\begin{tabular}{lcccc}
\toprule
direction & $U$ & full-collateral & ratio & effective support\\
\midrule
equal & $44.14$ & $1.04$ & $42.5\times$ & $100.0$\\
inverse $\sqrt{p(1-p)}$ & $57.04$ & $1.02$ & $55.8\times$ & $80.2$\\
inverse $p$ & $48.03$ & $1.01$ & $47.4\times$ & $54.5$\\
bucket-balanced & $44.14$ & $1.04$ & $42.5\times$ & $100.0$\\
\bottomrule
\end{tabular}
\end{table}

\begin{figure}[p]
\centering
\includegraphics[width=\textwidth]{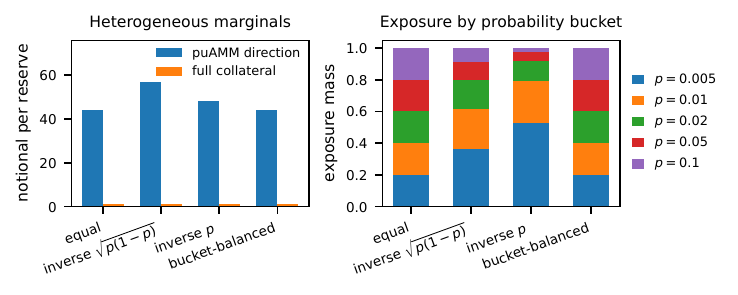}\\[4pt]
\includegraphics[width=\textwidth]{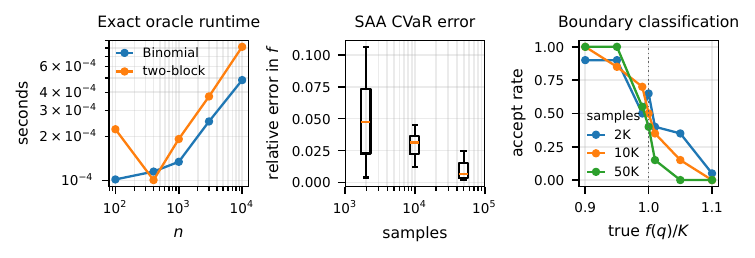}
\caption{Top: heterogeneous-market capacities and exposure mass by bucket.
Bottom: computation and sample-average boundary checks. Near $f(q)=K$, sampling noise can create both false accepts and false rejects, motivating safety
margins for approximate oracles.}
\label{fig:app-heterogeneous-scalability}

\vspace{14pt}
\includegraphics[width=\textwidth]{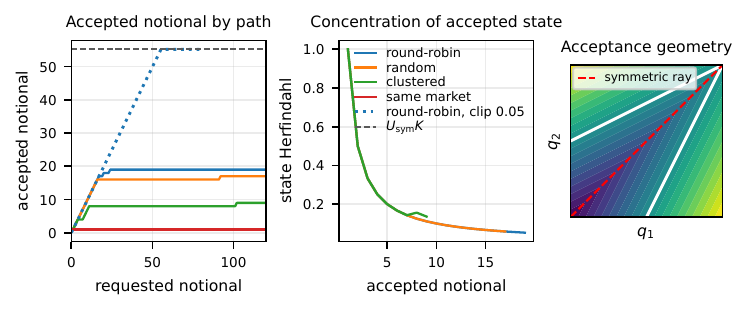}
\caption{Left and middle: exogenous order-arrival paths. Concentrated paths hit refusal early, while small diversified clips approach the symmetric capacity certificate.
Right: acceptance geometry for two markets; the white contour is
$f(q)=K$ and the dashed ray is the symmetric direction.}
\label{fig:app-order-geometry}
\end{figure}

\paragraph{Computational representation.}
Figure~\ref{fig:app-heterogeneous-scalability} (bottom) compares exact Binomial
evaluation, exact bucket convolution, and sample-average approximation. The
boundary experiment stresses a practical point: when the true state is close to
$f(q)=K$, finite sampling can misclassify the acceptance test. In the
$n=100$, $p=0.02$ symmetric case, sample-average tests with 50,000 scenarios
accept all states at true $f(q)/K\in\{0.90,0.95\}$ in the trial grid, but accept
only $55\%$ of states at $0.99$, $40\%$ at the boundary, and still falsely
accept $15\%$ at $1.01$. The practical takeaway: compute the acceptance test
off-chain from an oracle representation of the law of $q\cdot X$, with safety
margins near the boundary.

\paragraph{Order-arrival paths.}
Corollary~\ref{cor:capacity} bounds accepted notional on every path, and small trades approach the bound. Figure~\ref{fig:app-order-geometry} (left and middle) uses exogenous paths at $(n,p)=(100,0.02)$: with a round-robin clip of $0.05K$, accepted notional reaches $55.15K$, or $0.997\,U_{\mathrm{sym}}K$, as the approachability construction predicts, while coarse unit clips stop earlier, at $19K$ for round-robin, $17K$ for random, $9K$ for clustered, and $1K$ for same-market flow.
These are stress paths rather than demand predictions.

\paragraph{Canonical examples.}
Table~\ref{tab:app-examples} gives closed-form or exact finite examples for
the main regimes. A single Bernoulli claim and the comonotone endpoint both
have $U=2$ at $p=0.05$ because the two \ac{CVaR} levels see essentially the same
single tail event. The i.i.d.\ symmetric direction improves with diversified
tail averaging, reaching $U=39.26$ at $n=100$. Two independent comonotone
blocks sit between these extremes, with $U=8$. The small-$n$ rows are
discrete-tail examples and do not show the asymptotic
$\sqrt n$ regime.

\begin{table}[t]
\centering
\caption{Worked capacity examples at $p=0.05$.}
\label{tab:app-examples}
\begin{tabular}{lccc}
\toprule
example & $n$ & tail gap $f$ & certificate $U$\\
\midrule
single Bernoulli & $1$ & $0.5000$ & $2.00$\\
i.i.d.\ symmetric & $2$ & $0.1250$ & $8.00$\\
i.i.d.\ symmetric & $10$ & $0.1122$ & $8.91$\\
i.i.d.\ symmetric & $100$ & $0.0255$ & $39.26$\\
two independent comonotone blocks & $100$ & $0.1250$ & $8.00$\\
comonotone endpoint & $100$ & $0.5000$ & $2.00$\\
\bottomrule
\end{tabular}
\end{table}

\paragraph{TV bound in exact arithmetic.}
The bound of Theorem~\ref{thm:tv-robustness} is checkable in exact arithmetic
at small $n$. Take $n=10$, $p=0.05$, and the common-shock mixture
$(1-w)\,\mathrm{Bern}(p)^{\otimes n}+w\,\delta_{\text{all-YES}}$ for ten
weights $w\in[0,0.02]$, compared against the same-marginals product reference;
both laws and their \ac{TV} distance $\epsilon$ are enumerated exactly over the
$2^{10}$ outcomes. The realized symmetric-direction tail-gap error stays below
the bound $\epsilon(1/\sigma_1+1/\sigma_2)$ at every $w$ and is $10.8$ to $24.8$ times smaller across the grid, because the all-YES contamination moves both \ac{CVaR} levels together, so most of the shift
cancels in the gap $f$.
\end{document}